# Magnetic Fields Effects on the Electronic Conduction Properties of Molecular Ring Structures

Dhurba Rai, Oded Hod and Abraham Nitzan

School of Chemistry, Tel Aviv University, Tel Aviv 69978, Israel.

## Abstract

While mesoscopic conducting loops are sensitive to external magnetic fields, as is pronouncedly exemplified by observations of the Aharonov-Bohm (AB) effect in such structures, the small radius of molecular rings implies that the field needed to observe the AB periodicity is unrealistically large. In this paper we study the effect of magnetic field on electronic transport in molecular conduction junctions involving ring molecules, aiming to identify conditions where magnetic field dependence can be realistically observed. We consider electronic conduction of molecular ring structures modeled both within the tight-binding (Hückel) model and as continuous rings. We also show that much of the qualitative behavior of conduction in these models can be rationalized in terms of a much simpler junction model based on a two-state molecular bridge. Dephasing in these models is affected by two common tools: the Büttiker probe method and coherence damping within a density matrix formulation. We show that current through benzene ring can be controlled by moderate fields provided that several conditions are satisfied: (a) conduction must be dominated by degenerate (in the free molecule) molecular electronic resonances, associated with multiple pathways as is often the case with ring molecules; (b) molecular-leads electronic coupling must be weak so as to affect relatively distinct conduction resonances; (c) molecular binding to the leads must be asymmetric (e.g., for benzene, connection in the meta or ortho, but not para, configurations) and, (d) dephasing has to be small. When these conditions are satisfied, considerable sensitivity to an imposed magnetic field normal to the molecular ring plane is found in benzene and other aromatic molecules. Interestingly, in symmetric junctions (e.g. para connected benzene) the transmission coefficient can show sensitivity to magnetic field that is not reflected in the current-voltage characteristic. The analog of this behavior is also found in the continuous ring and the two level models.

Although sensitivity to magnetic field is suppressed by dephasing, quantitative estimates indicate that magnetic field control can be observed in suitable molecular conduction junctions.

# 1. Introduction

Controlling electron transmission through molecular junctions that comprise molecular ring structures by magnetic fields is considered challenging because the required field strengths are believed to be unrealistically high, of the order of the Aharonov-Bohm period of $\sim 10^4$ Tesla for typical molecular rings.[1, 2] In contrast, it was demonstrated theoretically[3-6] that the conductance of a nano-size ring can be significantly modulated by relatively moderate magnetic fields (< 50 Tesla). This large sensitivity to an external magnetic field results from the presence of sharp resonances that are possible only for low coupling between the molecular-bridge and the metal-contacts.

Recent studies of electronic conduction through molecular ring structures such as benzene, biphenyl, azulene, naphthalene and anthracene as well as carbon nanotubes, by us[7] and others[8-11] have shown that although the net current through these molecules at low metal-molecule coupling is relatively small, induced circular currents can be considerable. The magnitude of such a voltage driven ring current depends significantly on the metal-molecule coupling strength, while its very existence depends on junction geometry, specifically on the location and symmetry of the molecule-metal contact along the circumference of the ring. The magnetic field associated with such a circular current at the center of a molecular ring can be quite significant, for instance, we have found $\sim 0.23$ Tesla at 2 Volt bias in a tight-binding model of a meta-connected benzene bridge.[7] Possible exploitations of such high local magnetic fields at the molecular level could come through the realization of carbon nanotubes as molecular solenoids,[11-13] or by controlling the alignment of spin orientation of magnetic ions embedded in the bridge as suggested in Ref. [14].

Voltage driven circular currents in molecular ring structures are similar in nature to the persistent currents induced in isolated mesoscopic rings that are threaded by magnetic fluxes. Both phenomena originate from splitting of degenerate ring electronic states characterized by opposite orbital angular momenta, either by the magnetic fields or by the voltage bias even in the absence of external magnetic field.[15-19, 20] Theoretical studies have shown that magnetic flux induced persistent currents can be controlled by external radiation.[21, 22] Also, theory indicates that ring currents can be induced by polarized light,[23-25] twisted light,[26] and other optical coherent control methodologies,[27-29] Such optically induced circular currents can be effectively controlled by externally applied magnetic fields,[30] or, conversely, can be used to control the local magnetic field at the ring,[31-33] thereby opening the possibility to control the orientation of an impurity spin at the ring center.

In a recent preliminary study,[34] we have reconsidered the possibility of controlling electrical conduction characteristics of molecular ring structures by static uniform moderate magnetic fields, following the lead of Refs. [3-6], which indicate that large sensitivity to magnetic fields may be found in junctions where (a) electronic state degeneracy leads to interference that can be suitably tuned by the magnetic field, and (b) weak molecule-metal coupling results in sharp transmission resonances. We have shown that for certain geometries and under specified conditions magnetic field effects on molecular ring conduction can become observable. In the present study we analyze in detail the role of molecular geometry and dephasing on the transport properties of molecular rings such as benzene, biphenyl and anthracene, subject to external magnetic fields by means of tight-binding (Hückel) molecular ring models as well as a scattering matrix approach to transmission through continuum loops. The close similarity between the results obtained from these different models indicates their generic nature, and provides evidence to the integrity of results obtained for the effect of a magnetic field in the limited-basis tight binding model (the London approximation). In addition we show that the essential physics underlying the observed magnetic field dependence of conduction can be obtained already from a suitably constructed two-state model. The important role of interference processes implies that the system behavior will strongly depend on the junction geometry that determines the transmission pathways and on the effect of dephasing (decoherence) processes resulting from thermal motions in the junction. In this regard we note that an earlier study[35] indicates that increasing electron-vibration coupling in the junction may lead to sharper resonance structure in the current dependence on the magnetic field, and therefore to larger sensitivity of conduction to such field. This however is not simply related to pure dephasing as the effect discusses by Ref. [35] increases at lower temperatures.

The structure of the paper is as follows: in section 2, we present three different models for electron transmission under the influence of an external magnetic field: (i) A tight-binding (Hückel) model, which is analyzed using the steady state approach described in our earlier work[7,36,37] and (ii) continuum ring model analyzed using scattering theory[6]. We also describe (iii) a simple 2-level model that captures the essence of the observed behavior. Section 3 presents results from our model calculations that describe (a) the magnetic field effect on the transmission probability and the current voltage characteristics of simple molecular ring junctions of various symmetries; (b) comparison between the tight-binding, continuum and two-level models and (c) the effect of structure and geometry on the dependence of junction transport properties on the applied magnetic field. The effect of dephasing processes on these behaviors is discussed in Section 4. Section 5 summarizes our main results and discusses their implications for further study.

## 2. Models and Methods

In this section we describe the models used in this work to analyze magnetic field effects on electron transport through molecular rings. The tight-binding (Hückel) model seems to be the most suitable for a simple description of molecular transport, however we will see in the following sections that the main characteristics of the transport behavior are found also in the continuous ring model. In fact, much of the physics is already contained in an ever simpler model based on a two-level bridge. These models are described below.

**(a) The tight-Binding Model: Scattering theory on a 1-d lattice**

We consider a molecular junction formed by a ring molecule bridging the metal leads ($L$, $R$) through two chosen sites on the ring. The molecule is described by a tight-binding (Hückel) model with on-site energies $\alpha_M$ and nearest-neighbor interactions $\beta_M$. The metal contacts ($L$, $R$) are modeled as infinite 1-dimensional tight-binding periodic arrays of atoms with lattice constant $a$ and on-site energies and nearest-neighbor coupling matrix elements $\alpha_K$ and $\beta_K$ $(K \in L, R)$, respectively. A sketch of such a molecular junction is shown in Fig. 1

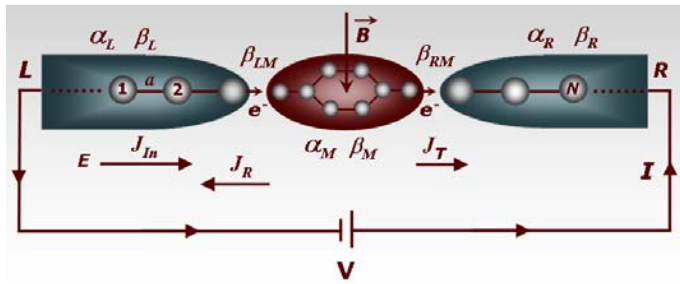

Fig. 1. A tight-binding model for current conduction through a molecular ring structure connected to two 1-dimensional metal leads $L$ and $R$, at bias voltage $V$. A static uniform magnetic field $\vec{B}$ is applied perpendicular to the molecular plane.

In the site representation, this TB Hamiltonian is given as

$$\hat{H} = \hat{H}_L + \hat{H}_R + \hat{H}_M + \hat{V}_{LM} + \hat{V}_{RM}, \qquad (1)$$

where

$$\hat{H}_K = \alpha_K \sum_{n \in K} |n\rangle\langle n| + \beta_K \sum_{n \in K} \left(|n\rangle\langle n+1| + |n+1\rangle\langle n|\right) ; \quad K = L, R, M \qquad (2)$$

$$\hat{V}_{LM} = \beta_{KM} \left(|n\rangle\langle m| + |m\rangle\langle n|\right) ; \quad n \in K = L, R ; m \in M , \qquad (3)$$

where $\{|n\rangle\}$ is an orthogonal set of atomic orbitals centered at atomic sites $n$ ($n \in K = L, R, M$). The indices $K = L, R$ and $M$ correspond to the subspaces of the left (L), right (R) and molecule (M), respectively.

The model (1)-(3) is supplemented by an additional magnetic field $\vec{B}$ applied to the junction, so that the kinetic energy operator of an electron is modified according to $(2m)^{-1}\hat{p}^2 \to (2m)^{-1}(\hat{p}+|e|\hat{A})^2$, where $m$ and $e$ are the electron mass and charge. We assume that the field is uniform, applied in the bridge region only and, for the planar ring-molecules considered, perpendicular to the molecular plane. In the standard London approximation[38] one (a) represents this field by a vector potential in the symmetric gauge, $\vec{A}(\vec{r}) = \frac{1}{2}\vec{B} \times \vec{r}$, and (b) modifies the finite basis of field-free atomic orbitals $|n\rangle = \chi_n(\vec{r})$, where $\vec{r}$ is measured from the center of atom $n$, so as to account for the phase difference between wavefunctions centered on different atomic sites,

$$\chi_n(\vec{r}) \to \tilde{\chi}_n(\vec{r}) = e^{-i(|e|/\hbar)\vec{A}_n \cdot \vec{r}} \chi_n(\vec{r}) \qquad (4)$$

where $\vec{A}_n = \vec{A}(\vec{r}_n)$. This leads to a tight binding (Hückel) Hamiltonian with coupling between atomic sites given by

$$\beta_{mn} = \int d\vec{r} \chi_m^*(\vec{r}) V \chi_n(\vec{r}) e^{-i(|e|/\hbar)(\vec{A}_n - \vec{A}_m) \cdot \vec{r}} \qquad (5)$$

For nearest neighbor interactions, $\vec{r}$ in the exponent is approximately replaced by $(\vec{r}_m + \vec{r}_n)/2$, leading to

$$\beta_M \to \beta_{mn} = \beta_M e^{i\theta_{mn}} ; \quad n, m \in M , \qquad (6)$$

where

$$\theta_{mn} = -\frac{|e|}{\hbar}(\vec{A}_n - \vec{A}_m) \cdot \frac{\vec{r}_m + \vec{r}_n}{2} = \frac{|e|}{4\hbar} \vec{B} \times (\vec{r}_m - \vec{r}_n) \cdot (\vec{r}_m + \vec{r}_n) \\ = \frac{|e|}{2\hbar} \vec{r}_m \times \vec{r}_n \cdot \vec{B} \qquad (7)$$

and the molecular Hamiltonian is now given by

$$\hat{H}_M = \alpha_M \sum_{n \in M} |n\rangle\langle n| + \beta_M \sum_{n,m \in M} \left( e^{i\theta_{mn}} |m\rangle\langle n| + e^{-i\theta_{mn}} |n\rangle\langle m| \right). \tag{8}$$

Note that $|\vec{r}_m \times \vec{r}_n|$ is twice the area of the triangle spanned by the vectors, so $\theta_{mn} = 2\pi \delta\phi_B/\phi_0$, where $\delta\phi_B$ is the magnetic flux through the triangle spanned by the position vectors $(\vec{r}_n, \vec{r}_m)$, and $\phi_0 = h/|e|$ is the flux quantum.

To evaluate the transmission coefficient associated with this setup one could use the non-equilibrium Green function method, but here we follow the scattering method used in Refs. [36] and [37] that gives an easier access to bond currents and, in its density matrix version, can be generalized to account (approximately) for dephasing processes. In this approach, a wire or a network of wires described by a tight-binding Hamiltonian is made to carry current by injecting electrons at one of its sites (source site). At the same time absorption is affected at the "end sites" of other wires by adding the self-energy term

$$\Sigma_K(E) = \frac{(E - \alpha_K)}{2} - \frac{i}{2}\sqrt{4\beta_K^2 - (E - \alpha_K)^2} \equiv \Lambda_K(E) - (i/2)\Gamma_K(E) \tag{9}$$

to the site energy. This "absorption" represents the infinite extent of the wire and makes it possible to describe dynamics in an infinite system by a finite system calculation. For electrons injected at energy $E$, The steady state wavefunctions are written in the site $(\{|n\rangle\})$ representation in the form

$$\Psi(E,t) = \sum_n C_n(E,t)|n\rangle = e^{-i(E/\hbar)t} \sum_n \bar{C}_n |n\rangle \tag{10}$$

The orbital coefficients or amplitudes $\bar{C}_n(E)$ obtained by solving the Schrödinger equation for steady state situation provide the particle current between any two nearest-neighbor sites $(n, m)$ on wire $K$ as

$$J_{nm}^K(E) = \frac{2\beta_K}{\hbar} \text{Im}(\bar{C}_n \bar{C}_m^*) \quad \forall \ (n,m) \in K, \tag{11}$$

which, in the presence of magnetic field (applied in the molecular region), becomes

$$J_{nm}^K(E) = \frac{2\beta_K}{\hbar} \text{Im}(e^{i\theta_{nm}} \bar{C}_n \bar{C}_m^*) \quad \forall \ (n,m) \in K = M. \tag{12}$$

The transmission probability through any exit segment or bond is calculated as the ratio of inter-site current (bond current) to the incoming particle current, i.e.,

$$T_{nm}^K(E) = \frac{J_{nm}^K(E)}{J_{In}(E)}. \tag{13}$$

Under certain conditions, a bond current in a biased molecular ring structure may exceed the net transport current – a signature that a circular current $I_c$ has developed in the ring. In a previous paper,[7] we have defined the circular current as the sole source of (bias induced) magnetic flux through the ring. For a ring comprising $n$ identical bonds, divided by nodes into $N$ segments of $n_j$ bonds ($\sum_{j=1}^{N} n_j = n$) on which the current has been determined to be $I_j$, the circular current in a ring is given by[7]

$$I_c = \frac{1}{n}\sum_j I_j\, n_j\,. \tag{14}$$

It was also useful to define the *circular transmission coefficient*[7] as the ratio of circular current to the incoming current as

$$\mathcal{T}_c(E) = \frac{J_c(E)}{J_{In}(E)}\,, \tag{15}$$

note that this number can be larger than 1.

For a finite bias voltage, the net bond current between any two nearest neighbor sites in a two-terminal junction is obtained from the Landauer formula

$$I_{mn}(V) = \frac{e}{\pi\hbar}\int_{-\infty}^{\infty} \mathcal{T}_{nm}(E)\ (f_L(E) - f_R(E))\ dE\,, \tag{16}$$

where $f_K(E)$ and $\mu_K$ ($K = L, R$) are the Fermi functions and chemical potential of the left and right leads, respectively. In the calculations reported below, unless otherwise stated, we have taken the leads temperature to be zero, assumed that the potential bias falls on the metal-molecule interfaces and considered symmetric potential drop, that is $\mu_L = E_F + eV/2$ and $\mu_R = E_F - eV/2$.

**(b) Scattering in the continuous ring model**

Next, we consider the continuous ring model, in which the junction comprises a one-dimensional (1D) conducting ring connecting between two 1D leads (Fig. 2a). The symmetry of the scattering process is imposed through the angle $\gamma$ between the leads,[39] and the properties of the ring-lead contact are embedded in the imposed junction scattering amplitudes (Fig. 2b) as detailed below.

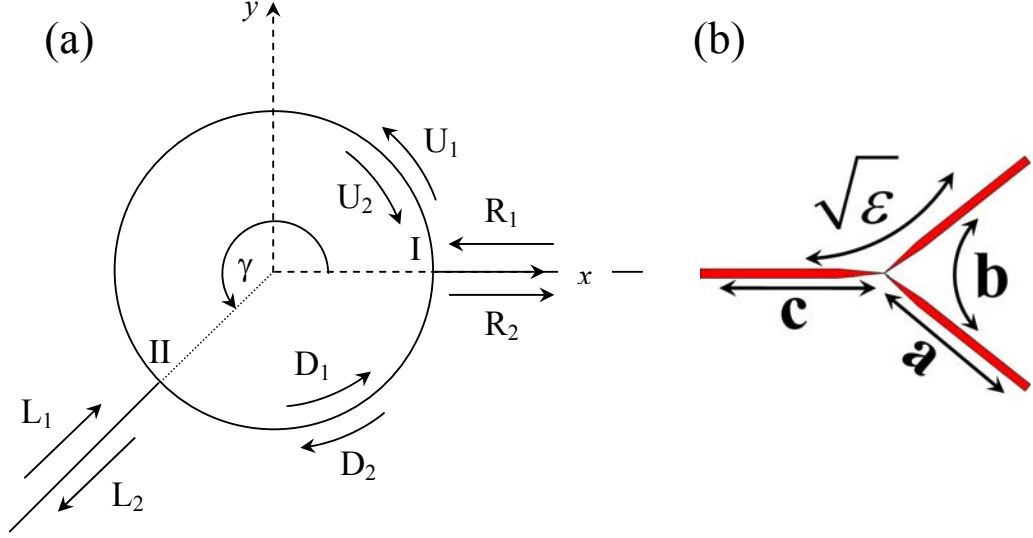

Fig. 2: Schematic representation of the scattering model. Panel (a): assignment of the wave amplitudes on the different parts of the system. Panel (b): junction scattering amplitudes.

We assume that the electrons can be represented by plane waves traveling along the ring with the form $A_1 e^{ikl} + A_2 e^{-ikl}$ where $l$ is the electron path length (on the ring $l = R\theta$, where $\theta$ is the angle traversed by the electron, and $R$ – the ring radius) and $A_{1,2}$ are the amplitudes of the respective waves (In Fig. 2a these amplitudes are denoted $L$, $U$, $D$ and $R$ in different segments of the ring and the leads). We use the standard notation where negative(positive) $k$ values represent (counter-)clockwise propagating waves. The ring-lead coupling is modeled by assigning scattering amplitudes at the ring-leads junctions[40-46] as shown in Fig. 2b. $|c|^2$ is the probability of an electron approaching the junction from the lead to be back scattered into the lead, $\varepsilon$ is the probability of an electron approaching the junction from the lead to mount the ring, $|a|^2$ is the probability of an electron approaching the junction from one of the arms of the ring to be back scattered into the same arm, and $|b|^2$ is the probability of an electron approaching the junction from one of the arms of the ring to be transmitted into the other arm. Based on current conservation considerations the scattering matrix can be shown to be unitary such that all scattering amplitudes (taken to be real[42, 47, 48]) can be folded into a single parameter which we choose to be $\varepsilon$ such that:

$$c = \sqrt{1-2\varepsilon} \quad ; \quad a = \frac{1}{2}(1-c) \quad ; \quad b = -\frac{1}{2}(1+c) \qquad (17)$$

It is now possible (Appendix A) to write scattering equations for both junctions taking into account the spatial and magnetic phases accumulated by the electrons while traveling along the arms of the ring. Focusing on the scattering process associated with an incident wave coming on the right lead with amplitude $R_1$, that is, taking the incoming amplitude on the left lead to vanish, $L_1=0$, these equations can be used to relate all amplitudes in the ring segment and the leads to the incoming amplitude $R_1$. This leads (see Appendix A) to the transmission probability in the form

$$T(k,\phi_B) = \left|\frac{L_2}{R_1}\right|^2 = \frac{\mathcal{T}_1}{\mathcal{T}_2} \tag{18a}$$

$$\mathcal{T}_1 = 4\varepsilon^2\left\{1-\cos(2\pi kR)\cos[2kR(\pi-\gamma)]+2\sin[kR(2\pi-\gamma)]\sin(kR\gamma)\cos\left(2\pi\frac{\phi_B}{\phi_0}\right)\right\} \tag{18b}$$

$$\mathcal{T}_2 = (c^2-1)^2 + 4\left\{\cos(2\pi kR)-a^2\cos[2kR(\pi-\gamma)]-b^2\cos\left(2\pi\frac{\phi_B}{\phi_0}\right)\right\}$$
$$\times\left\{C^2\cos(2\pi kR)-a^2\cos[2kR(\pi-\gamma)]-b^2\cos\left(2\pi\frac{\phi_B}{\phi_0}\right)\right\} \tag{18c}$$

where $\phi_B = \vec{B}\cdot\vec{S} = B_z S$, $\vec{S}$ is a vector perpendicular to the ring's surface such that $S = |\vec{S}| = \pi R^2$ and, as above, $\phi_0 = 2\pi\hbar/|e|$. We note that when $\gamma = \pi$ the above expression reduces to the standard expression for symmetrically connected rings.[49]

**(c) A two State Model**

In the weak leads-ring coupling limit, the width of the doubly-degenerate energy levels on the ring is considerably smaller than the inter-level spacing between states of different angular momentum. Therefore, at low bias voltages, we can safely assume that electronic transport takes place mainly through a couple of degenerate levels close to the Fermi energy of the leads and model the transport physics using the simplified two-level model shown in Fig. 3. To assign the relevant model parameters, consider an external magnetic field $\vec{B} = (0,0,B_z)$ threading a molecular ring of radius $R$ that lies in the XY plane. The energy levels of an electron moving otherwise freely on the ring are given by (see Appendix B):

$$E_m = \frac{1}{2R^2}\left(m + \frac{\phi_B}{\phi_0}\right)^2 \tag{19}$$

where we use atomic units unless otherwise stated. Here, $m = 0, \pm 1, \pm 2, \cdots$ is the angular quantum number (for the isolated ring, the quantum number $m$ relates to the wave vector $k$ of the previous section via $m = kR$) and the flux quantum is $\phi_0 = 2\pi$. In the basis of the corresponding eigenstates of the isolated ring, the molecular ring Hamiltonian $\hat{H}_M$ is given by a repeated sequence of diagonal 2x2 blocks, $\hat{H}_M^{(m)}$, that correspond to states whose degeneracy is split by the field:

$$\hat{H}_M^{(m)} = \begin{pmatrix} \frac{1}{2R^2}\left[m^2 + \left(\frac{\phi_B}{\phi_0}\right)^2\right] - \frac{|m|B_z}{2} & 0 \\ 0 & \frac{1}{2R^2}\left[m^2 + \left(\frac{\phi_B}{\phi_0}\right)^2\right] + \frac{|m|B_z}{2} \end{pmatrix} \equiv \begin{pmatrix} E_1 & 0 \\ 0 & E_2 \end{pmatrix}. \tag{20}$$

The full transport problem of a molecular ring connecting between two metal leads can thus be replaced by the simplified model involving only the two levels characterized by a given $|m|$, shown in Fig. 3. Here $V_{j,K}$; $j = 1,2$; $K = L, R$ denotes the coupling between molecular level $j$ and the lead $K$, and

$$E_{1/2} \equiv \frac{1}{2R^2}\left[m^2 + \left(\frac{\phi_B}{\phi_0}\right)^2\right] \mp \frac{|m|B_z}{2} \tag{21}$$

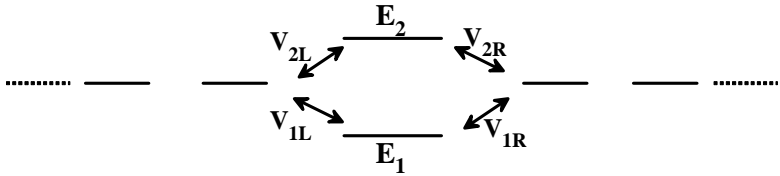

Figure 3: Schematic representation of the two-level model. The leads are represented in this figure by tight binding chains.

The corresponding transmission coefficient is given by the Landauer formula

$$\mathcal{T}(E) = Tr\left[\Gamma_L(E) G_M^r(E) \Gamma_R(E) G_M^a(E)\right] \tag{22}$$

where the broadening function of lead $K$ (=L,R) is given by

$$\Gamma_K(E) = i\left[\Sigma_K^r(E) - \Sigma_K^a(E)\right] \tag{23a}$$

the retarded Greens function of the molecule is calculated as

$$G_M^r(E) = \left[G_M^a(E)\right]^\dagger = \left[EI - H_M - \Sigma_L^r(E) - \Sigma_R^r(E)\right]^{-1} \quad (23b)$$

And the self-energy of lead $K$ is represented as

$$\Sigma_K^r(E) = \left[\Sigma_K^a(E)\right]^\dagger = \left[G_K^{r0}(E)\right]_{kk} \begin{pmatrix} |V_{1,K}|^2 & V_{1,K}^* V_{2,K} \\ V_{2,K}^* V_{1,K} & |V_{2,K}|^2 \end{pmatrix} \quad (23c)$$

$G_K^{r0}(E)$ being the retarded Green's function of the isolated lead $K$. Eq. (23c) is written under the assumption of short range interaction between ring and lead, whereby the ring is coupled to the nearest neighbor lead site denoted by the index $k$. We could have used here the explicit Newns-Anderson model for a 1-dimensional tight binding lead for which $\Sigma$ is given by Eqs. (9), however, because we will be using the 2-level model as a generic simple model to gain physical insight into the nature of our results it is enough to make the simplest wide-band approximation for which $\left[G_L^{r0}(E)\right]_{kk} = \left[G_R^{r0}(E)\right]_{kk} \approx -i\pi\rho$ where $\rho$ is the density of lead electronic states and we assume identical leads. On the other hand, the magnetic field dependent transport properties are determined by the choice of lead-ring coupling elements that enter Eq. (23c). Aiming to capture this dependence, we assume that $\langle \psi_{ring} | \hat{V} | \psi_{lead} \rangle$ reflects the phase of the wave function on the ring at the positions of the leads-ring junction. Referring to Fig. 2a and setting the angle at which the right lead is attached to the ring to be $\gamma_0 = 0$, the left lead is attached at the respective angle $\gamma$ ($\gamma$ is $\pi$, $2\pi/3$ and $\pi/3$ for the para, meta and ortho configurations, respectively). Hence we take

$$V_{1,R} = Ve^{-im\gamma_0} = V; \quad V_{2,R} = Ve^{im\gamma_0} = V; \quad V_{1,L} = Ve^{-im\gamma}; \quad V_{2,L} = Ve^{im\gamma} \quad (24)$$

where $V$ is the coupling strength. Consequently, Eq. (23c) yields the retarded self-energies associated with the right and left leads in the forms

$$\Sigma_R^r(E) = -i\pi\rho V^2 \begin{pmatrix} 1 & 1 \\ 1 & 1 \end{pmatrix}; \quad \Sigma_L^r(E) = -i\pi\rho V^2 \begin{pmatrix} 1 & e^{2im\gamma} \\ e^{-2im\gamma} & 1 \end{pmatrix} \quad (25)$$

where $m$ is related to the imposed magnetic field and to the electron energy through Eq.(21). It is now possible to obtain explicit expressions for the broadening matrices $\Gamma_{L/R}(E)$ and the retarded and advanced molecular Green's functions $G_M^r(E), G_M^a(E)$ from which the transmission probability can be calculated using Eq. (22). A long but straightforward calculation (see Appendix B for a detailed derivation) leads to

$$T(E, B_z) = \frac{\mathcal{T}_1}{\mathcal{T}_2} \tag{26a}$$

$$\mathcal{T}_1 = \Gamma^2 \left[ (E-E_1)^2 + 2(E-E_1)(E-E_2)\cos(2m\gamma) + (E-E_2)^2 \right] \tag{26b}$$

$$\mathcal{T}_2 = (E-E_1)^2 (E-E_2)^2$$
$$+ \Gamma^2 \left[ (E-E_1)^2 + 2(E-E_1)(E-E_2)\cos^2(m\gamma) + (E-E_2)^2 \right] + \left[ \Gamma \sin(m\gamma) \right]^4 \tag{26c}$$

$$\Gamma = 2\pi V^2 \rho \tag{26d}$$

where the dependence on $B_z$ originates from Eq.(21).

## 3. Results and discussion

As discussed above, current conduction through ring structures is inherently associated with interfering transmission pathways[50] that may be conveniently described in terms of degenerate eigenstates of the isolated ring. The corresponding degenerate states can be represented in terms of rotating, clockwise and counter-clockwise, Bloch states on the ring. Indeed, it is the tuning of relative phases of these states by magnetic field that potentially provides magnetic field control of the ring transmission properties. This implies several important aspects of the resulting behavior: First, transport will be affected by interference (and consequently most amenable to magnetic field control) in energy regimes dominated by such degenerate states. Second, strong interference effects and large sensitivity to magnetic field are expected when these states are associated with sharp transmission resonances, i.e., for sufficiently weak metal-molecule coupling. Third, the symmetry of a given junction geometry strongly affects the interference pattern and hence dictates the transport properties. Fourth, these phenomena will be strongly affected by dephasing processes. In what follows we will see different manifestations of these statements.

We study single-molecule junctions consisting of molecular ring structures such as benzene, biphenyl and anthracene connecting metal leads. The results presented below focus on the response of these systems to an externally applied static uniform magnetic field in terms of modification in their electronic transport characteristics. The molecular junction is emulated by the tight-binding (Hückel) molecular Hamiltonian and 1-dimensional tight binding leads as presented above. For the latter we take zero on-site energies, i.e., $\alpha_K$ $(K \in L, R) = 0$ and nearest neighbor coupling $\beta_L = \beta_R = 6$ eV that corresponds to a metallic band of width 24 eV. The zero bias Fermi energies of these contacts are set to $E_F = 0$. For the molecular structure we take $\alpha_M = -1.5$ eV and $\beta_M = 2.5$ eV for all nearest-neighbor atom pairs.[51]

Consider first a simple benzene ring that can couple to the metal leads in para, meta and ortho configurations (Fig. 1). In the free molecule the highest occupied molecular orbitals (HOMOs) and the lowest unoccupied molecular orbitals (LUMOs) constitute pairs of doubly degenerate orbitals which, with our choice of molecular parameters and energy origin, are positioned at $\alpha_M - \beta_M = -4\ eV$ and $\alpha_M + \beta_M = 1\ eV$, respectively. Upon connecting to the metal leads these levels get broadened and, more importantly, their degeneracy split. For sufficiently weak metal-molecule coupling these split levels constitute sharp transmission resonances at the corresponding energies.

It should be emphasized that degeneracy splitting in benzene affected by imposing perturbations at some atomic positions does not by itself specify the nature of the new eigenstates. An important property of the resonances obtained when the ring is connected to infinite leads at the meta or ortho positions (i.e. scattering resonances characterized by scattering boundary conditions, that is, incoming in one lead and outgoing in the other(s)), is that they can be shown to maintain the character of circulating Bloch eigenstates of the isolated ring. The corresponding transmission resonances are therefore associated with considerable circular current in the benzene ring.[7] Consequently, in the meta and ortho-connected configurations, large circular currents are found when bias and gate potentials are such that one of the split resonances dominates. In contrast, in the para connected ring, one of the split eigenstates turns out to have a node at one of the para positions and, consequently, does not contribute to transmission, while the other is characterized by zero net circular current as could be expected from symmetry.

This is shown in the upper panel of Fig. 4, where the bias voltage is set to $V = 2$ V. This brings the upper Fermi energy to the vicinity of the LUMO pair: In the meta and para connected benzenes one of the split resonances is below and the other above this energy.[52] The metal-molecule coupling is taken $\beta_{LM} = \beta_{RM} = 0.05$ eV, low enough so these resonances remain well separated. For these parameters the net current through the junction is of order ~ $nA$, while the circular current in both the meta and ortho configurations is three orders of magnitudes larger, yielding $\approx 0.23$ Tesla for the induced magnetic field in the ring center in both cases. Note that the direction of the circular current and the ensuing magnetic field is opposite in the meta and ortho configurations. See Ref.[7] for more details.

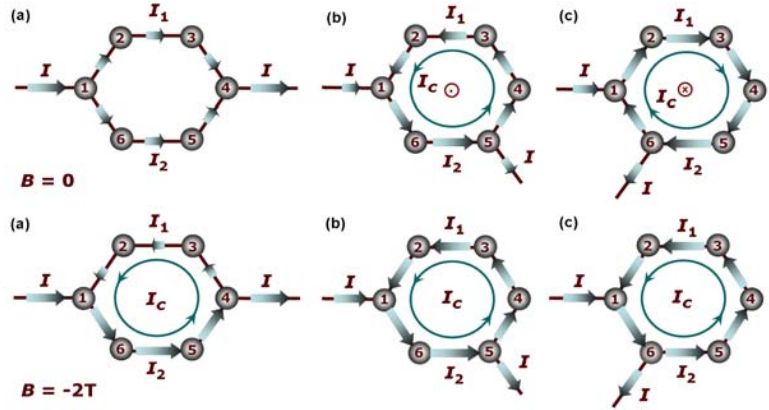

Fig. 4. Internal current distribution in (a) para (b) meta and (c) ortho-connected benzene ring, connected to leads with $\beta_{KM} = 0.05\,\text{eV}$ under voltage bias of $V = 2$ V. The upper panel shows the bond currents calculated in the absence of an external magnetic field, while the lower panel corresponds to the presence of a magnetic field, $B = -2\text{T}$ (negative $B$ corresponds to a field pointing down into the plane). The arrows along bonds represent bond currents with magnitudes proportional to the corresponding arrows lengths. The encircled dot (cross) in the meta (ortho) structures in the upper panel denote the directions of the magnetic field induced by the circular current: out of (into) the molecular plane.

When an external magnetic field is switched on, the bond-current map changes. In these and the following calculations the external magnetic field is taken in the Z direction, perpendicular to the molecular XY (also page) plane. Positive field direction is taken to be outwards, towards the reader and a positive circular current is taken to be in the counterclockwise direction. This field generates an additional circular, so called persistent, current that can reinforce or suppress the voltage driven circular current. Thus, a negative magnetic field (direction into the paper plane) generates a current in the anticlockwise direction that adds to the circular current in the meta connected ring (at $V = 2\,\text{V}$) and subtract from it in the ortho-connected structure, as seen in the lower panels of Fig. 4. Of course, the interplay between the voltage driven and field induced circular current depends on the voltage range considered. For example, in the meta-connected ring the voltage driven circular current reverses its direction above $V = 2\,\text{V}$ and a negative field induced persistent current will add to it destructively as in the ortho case. This implies that at any finite bias $I_c(B) \neq I_c(-B)$.

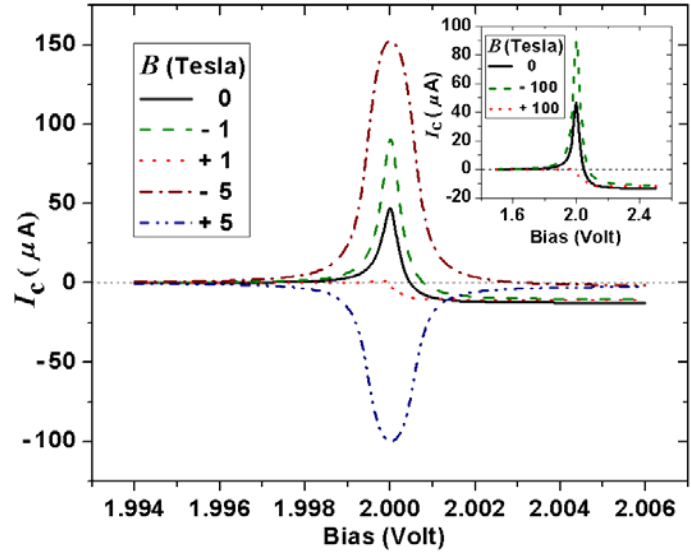

Fig. 5. Circular current as a function of bias voltage in the range (1.994 to 2.006 Volt) in a meta-connected benzene for $\beta_{KM} = 0.05\,\text{eV}$ in the presence of external magnetic field, $B = 0, +/- 1\text{T}$ and $+/- 5\text{T}$. The inset depicts the case for $\beta_{KM} = 0.5\,\text{eV}$ for applied field $B = 0, +/- 100\text{ T}$.

Fig. 5 shows another aspect of this effect. Here the circular current in a meta-connected benzene ring connected to leads with $\beta_{KM} = 0.05, 0.5\,\text{eV}$ ($K = L, R$) is shown as a function of voltage for different applied magnetic fields. Again it is seen that $I_c(B) \neq I_c(-B)$ in the presence of bias voltage. It is also seen (inset) that the sensitivity to magnetic field is strongly reduced when the molecule-lead coupling becomes stronger.

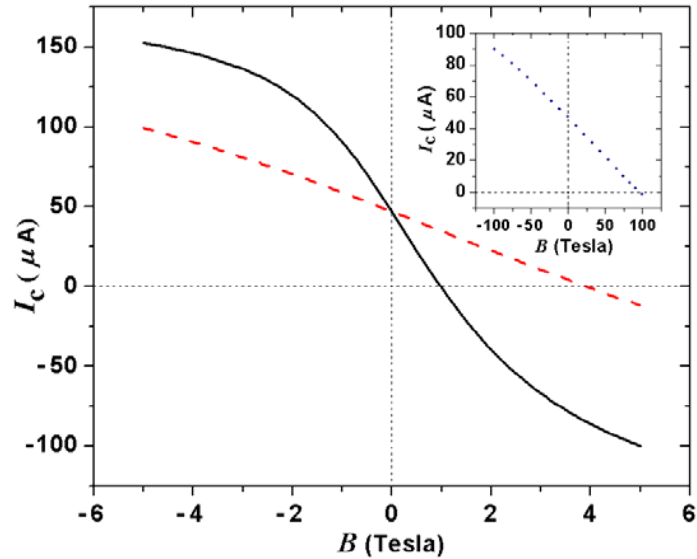

Fig. 6. The circular current in a meta-connected benzene ring biased at $V = 2$ V plotted as function of magnetic field applied perpendicular to the ring. The three cases shown correspond to different molecule-lead coupling. Full line (black): $\beta_{KM} = 0.05$ eV, dashed line (red): $\beta_{KM} = 0.10$ eV and dotted line (blue): $\beta_{KM} = 0.5$ eV (in the inset). The magnetic field induced by the voltage driven current is practically the same in all cases, $B_{ind} \simeq 0.23$ T.

It is of interest to ask, what is the external magnetic field that will annihilate the circular current in a voltage driven molecule? One may naively expect that this is just the field equal in magnitude and opposite in direction to that induced by the circular current so that the two fields annihilate each other, however Fig. 6 shows that the magnetic field needed to stop the circular current is considerably larger than the magnetic field produced by that current, and generally depends on the molecule-lead coupling. For $\beta_{KM} = 0.05, 0.1, 0.5$ eV we find this field to be 0.98, 3.92 and 96.40 tesla, respectively, at the voltage bias employed (2 V). On the other hand we find that the magnetic field induced by the circular current, $B_{ind} \cong 0.23$ tesla at the same voltage bias, almost independent of the molecule-lead coupling.[53] This non-trivial behavior results from the fact that the application of the external magnetic field does not simply oppose the circular-current induced magnetic field but also alters the electronic structure of the ring and strongly influences the interference pattern of coherent electrons mounting the ring thus influencing the resulting induced magnetic field itself.

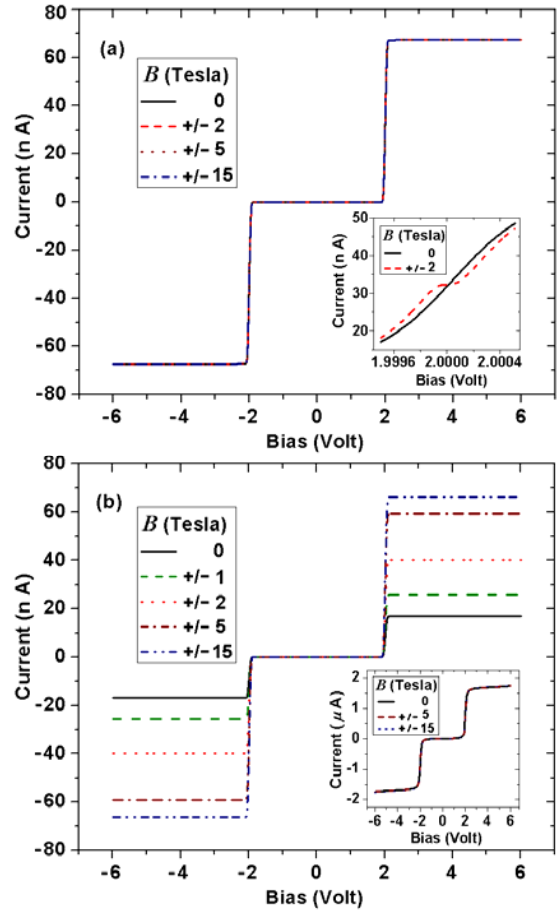

Fig. 7 The I-V Characteristics of a junction comprising para- (upper panel) and meta- (lower panel) connected benzene coupled to the leads with coupling element 0.05 eV, evaluated for different magnetic field strengths $B$ normal to the ring. The results obtained for different magnetic fields for the para system are essentially indistinguishable from each other. The inset in the upper panel shows a close-up on the $V = 2$ V neighborhood that shows the consequence of the split degeneracy in the para-connected junction. The inset in the lower panel shows the I-V behavior in the meta configuration for molecule-metal coupling 0.5 eV, which is essentially field independent in the same range of magnetic field strengths. Results for the ortho-connected molecule are qualitatively similar to those shown for the meta configuration.

Of more practical implications is the question whether the junction transport properties can be affected by an externally applied magnetic field. As already mentioned, previous studies[1,2] seem to indicate that while the transmission $\mathcal{T}(E)$ may be affected by an external magnetic field, the integrated transmission that yields current-voltage characteristics, is not sensitive to this field. The main reason for this observation is that at realistic magnetic field values the splitting between the degenerate levels of the ring is of the order of meVs (see Fig. 9). At high leads-ring coupling this splitting is much smaller than the width of the two levels and is therefore hardly seen even in the $T(E)$ curves. At the low coupling limit, the $T(E)$ curve clearly shows the magnetic field induced level splitting but in order to observe the magnetic field effect in the *I(V)* curves, the Fermi integration window (see Eq. (16)) should include only one of the split levels. This, however, requires bias and gate voltage precision smaller than the level splitting, as well as very low temperatures.

One may conclude that despite the ability to control the magnetic field sensitivity of the transmission probability through molecular rings via the leads-ring coupling,[5, 6] practical measurements of the *I(V)* curves will hardly show any magnetic-field effect. This can be clearly seen in the upper panel of Fig. 7 where the current-voltage relationship of a symmetrically (para-) connected benzene ring is found to be robust against the external field. Only when zooming into the current step region (inset of the upper panel of Fig. 7) one finds a small shoulder resulting from the level splitting at a finite magnetic field value.

While this conclusion is true for the symmetric junction, in the asymmetrically connected junction a different behavior is observed. In the lower panel of Fig. 7 we present the *I(V)* curves of the meta-connected ring for different magnetic field intensities. Despite the fact that the level splitting is similar to that of the symmetric junction, the current-voltage characteristics show pronounced sensitivity towards the magnetic field.

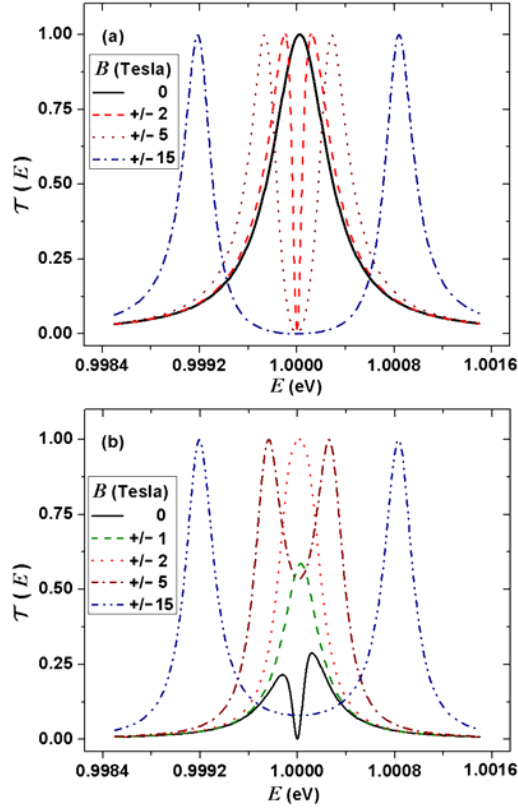

Fig. 8 Transmission probability $\mathcal{T}(E)$ around $E=1$ eV through a junction comprising para- (upper panel) and meta- (lower panel) connected benzene coupled to the leads with coupling matrix element 0.05 eV, evaluated for different magnetic field strengths $B$ normal to the plane of the ring. Again, results for the ortho-connected molecule are qualitatively similar to those shown for the meta configuration.

To get further insight into the origin of this behavior, we show in Fig. 8 the transmission functions $\mathcal{T}(E)$ that constitute the input to the I-V results of the preceding figure. Shown are the transmission functions for para- (upper panel) and meta- (lower panel) connected benzene rings under different perpendicular magnetic fields in the vicinity of the doubly degenerate LUMO energy of the isolated molecule, $\alpha_M + \beta_M = 1\ eV$. We see that the transmission function depends on the magnetic field in both the symmetric and the asymmetric junctions. Consider first the para connected junction and denote the split states in this configurations by $\psi_1$ and $\psi_2$. As noted above, only one of these, say $\psi_1$, contributes to the transmission and the system is characterized by a single transmission resonance. In terms of the two counter-rotating Bloch states that are eigenfunctions of the isolated ring, this resonance is a linear superposition in which the corresponding paths add constructively. The other state $\psi_2$ of zero transmission corresponds to the superposition in which they interfere destructively to give a node in one of the para positions. In the presence of an applied field $B$ the eigenstates become (as B→0) $2^{-1/2}(\psi_1 \pm \psi_2)$, implying that (a) the single transmission peak at $B=0$ splits into two peaks of equal intensities, and (b) the total area under the transmission function remains unchanged. This leads to an I-V characteristics that does not depend on the magnetic field (Fig. 7a) except in the very narrow voltage region where the Fermi step goes through the split peak (Fig. 7a inset). As discussed above, such a sharp Fermi step requires very low temperatures (~1 K) to be resolved.

In contrast, in the meta- and ortho- connected junctions, the asymmetric coupling to the leads results in the appearance of two transmission peaks to appear already in the absence of external magnetic field. As $B$ is increased, this splitting reduces up to a certain magnetic field intensity (for instance, +/-2 tesla in the meta-configuration for 0.05 eV molecule-lead coupling) where it vanishes (level crossing) engendering constructive interference at this field value. This can be understood as phase adjustment of the interfering electron waves by field, causing them to interfere constructively until full resonant transmission is reached. Interestingly, in this regime of magnetic field strength not only the splitting but also the area under the transmission function is field dependent. As the field intensity increases from zero the total area under the peaks increases until the peaks become fully separated and then the area remains constant. This is clearly manifested in the field dependence of the current-voltage characteristic shown in Fig. 7b, suggesting that in the asymmetric case the I-V field dependence should be experimentally accessible in the low leads-ring coupling regime. Note that Fig. 7b shows results obtained at 0K, however the results obtained at 300K are almost indistinguishable.

A more general view of this behavior is seen in Fig. 9, which shows, for the para- and meta- connected junctions, the evolution of the benzene energy levels at $E = \alpha_M + \beta_M = 1\ eV$ as a function of molecule-leads coupling (left side of figures) and magnetic field (right) as expressed by the transmission function. It is seen that level degeneracy is lifted by the molecule lead coupling in the meta structure, but not in the para structure. Increasing the magnetic field for a given (0.05 eV) molecule-lead coupling splits the levels in the para case, but brings them together first in the meta case, as discussed above.

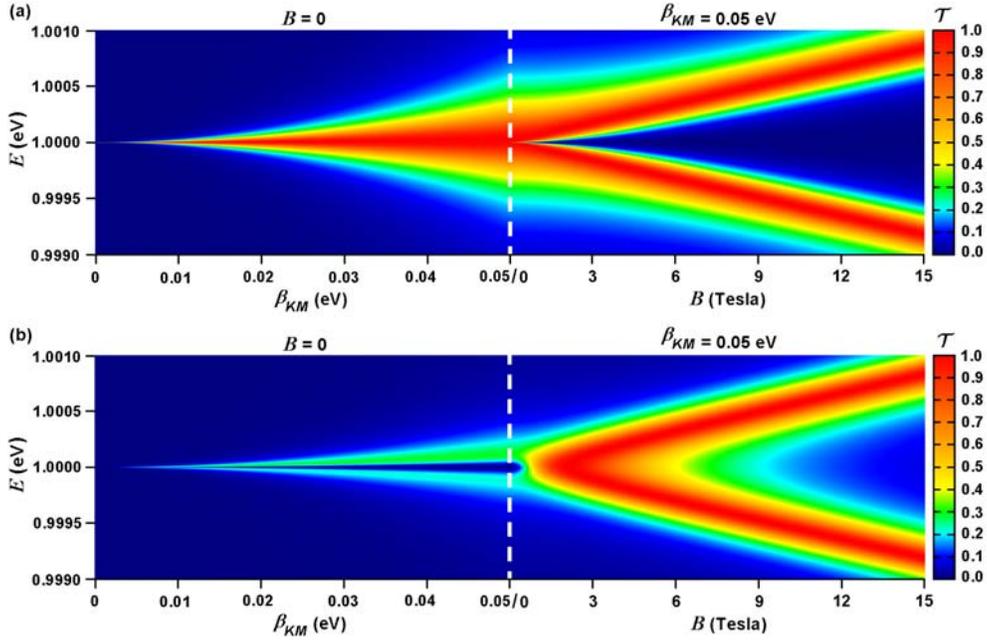

Fig. 9. Transmission probability maps in the $E - \beta_{KM}$ $(K = L, R)$ plane at $B = 0$ (left of the vertical dashed line) and in the $E - B$ plane for $\beta_{KM} = 0.05\,\text{eV}$ (right of the vertical dashed line) for junctions comprising para- (upper panel) and meta-connected (lower panel) benzene molecules. Color code varies from deep blue ($\mathcal{T} = 0$) to red ($\mathcal{T} = 1$).

Two additional observations should be pointed out. First, another regime of field dependence takes place at very high fields, where shift of energy levels makes more levels to appear within the Fermi window between $\mu_L$ and $\mu_R$. This happened at unrealistically large fields $(\sim 1000\text{T})$. Second, in contrast to the circular transmission coefficient $\mathcal{T}_c(E)$, the total transmission coefficient $\mathcal{T}(E)$ is not affected by the field direction, $I(B) = I(-B)$.

Qualitatively similar results as described above are obtained for other ring-containing molecular systems. For example, the isolated biphenyl molecule, which comprises two coupled benzene rings, possesses two 2-fold degenerate orbitals, viz., HOMO-1 and LUMO+1 that for our choice of parameters are positioned at $\alpha_M - \beta_M = -4\ eV$ and $\alpha_M + \beta_M = 1\ eV$, respectively. A biphenyl molecule connected to leads at positions 6,10 (see Fig. 10) can be considered as two coupled benzene rings in para configuration, and consequently we expect that its I-V characteristic is insensitive to an imposed weak magnetic field. Indeed, a recent work[54] finds that to affect transport in this configuration by magnetic field requires flux of order $0.5\phi_0$ which, as discussed, is unrealistic for such a small molecular structure. On the other hand, the diagonally connected biphenyl shown in Fig. 10a can be considered as two coupled benzene rings, each in meta configuration. For such a structure weakly coupled to leads, we find again high sensitivity of the transmission (Fig. 10b) and the I-V curve (Fig. 10c) to the magnetic field. It should be noted that sensitivity to magnetic field is manifested when the molecular levels at 1eV (degenerate in the free molecule) enter the Fermi window at voltage bias 2V. The current rise at V = 0.52V and 3.58V is due to resonant transmission through non-degenerate energy levels at ~ 0.26 eV (LUMO) and 1.79 eV (LUMO+2), respectively, and is not affected by the field.

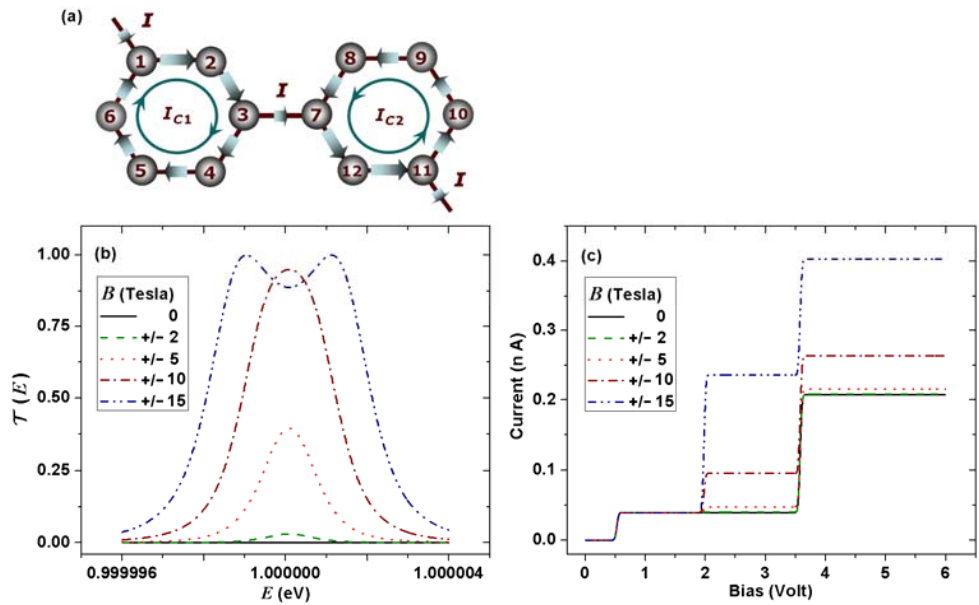

Fig. 10. (a) Field free internal current distribution in a diagonally connected biphenyl molecule at a bias voltage of 2V showing the circular currents in the absence of external magnetic field for metal-molecule coupling of 0.005 eV. (b) The transmission probability displayed against the electron energy around 1 eV at different field strengths. (c) Magnetic field effects on the I-V characteristic for $|\vec{B}|$ in the range 0…15 Tesla.

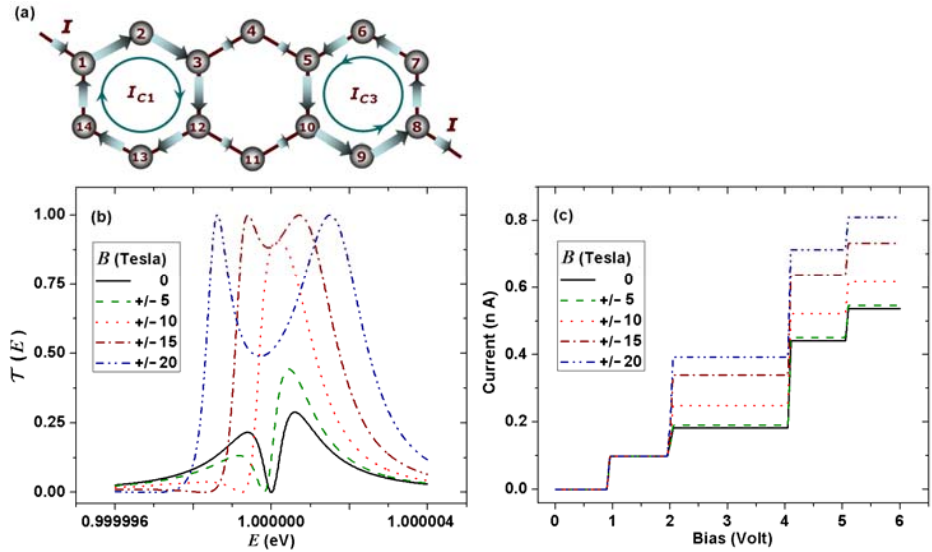

Fig. 11. (a) Field free internal current distribution in a diagonally connected anthracene molecule at a bias voltage of 2V showing the circular currents in absence of external magnetic field for metal-molecule coupling of 0.005 eV. (b) The transmission probability displayed against the electron energy around 1 eV at different magnetic field strengths. (c) Magnetic field effects on the I-V characteristic for $|\vec{B}| \leq 20$ Tesla.

Similar results for a junction with diagonally connected anthracene bridge are shown in Fig. 11. In the voltage range shown, sensitivity to a weak magnetic field is associated with the doubly degenerate anthracene levels at $\alpha_M + \beta_M = 1\ eV$ (LUMO+1) and $\alpha_M + \sqrt{2}\beta_M \approx 2.03\ eV$ (LUMO+2) (responsible for the current rise at $V$ = 2 and 4.06 eV, respectively). The other current steps seen in Fig. 11c are due to non-degenerate levels and are not sensitive to the field. As in the previous cases discussed, this behavior depends on the symmetry of the molecular junction. For example, in agreement with Ref. [55], no sensitivity of the I-V characteristic to weak fields is found for contacts placed in the (2,6) positions (see Fig. 11a) although the transmission function itself does depend on the field in a way reminiscent to the para-connected benzene junction.

Finally, we note that naphthalene does not have orbital degeneracy and indeed no I-V sensitivity to weak magnetic field is found (although circular currents are induced) in junction models based on this structure. Conduction through this molecule is found to be affected only by unrealistic strong fields of order ~ 1000 tesla. These observations can be summarized by stating that sensitivity of the I-V behavior to relatively weak external magnetic fields is a generic phenomenon in many ring molecular structures characterized by weak molecule-lead coupling, however its manifestation depends on details of the electronic structure of the molecule and on the junction geometry.

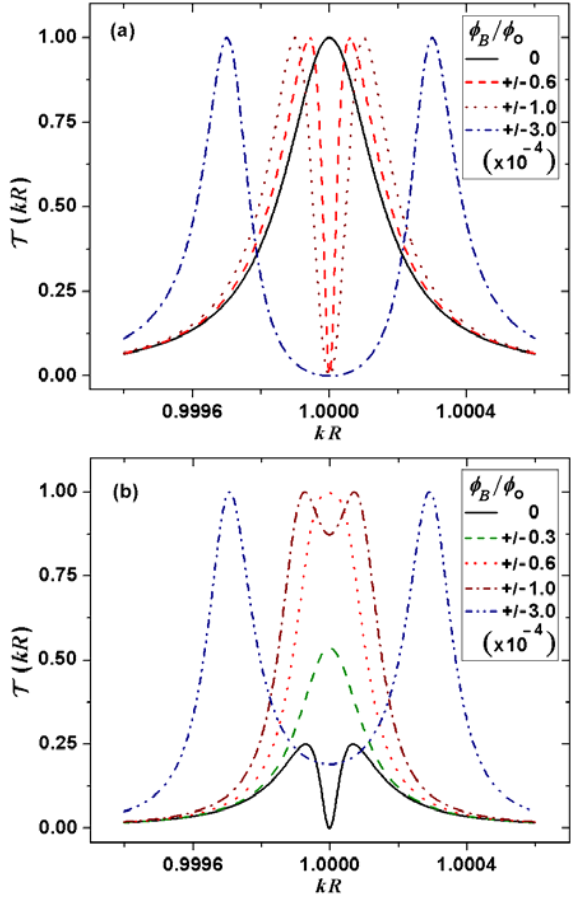

Fig. 12. Transmission probability around $kR = 1$ in a continuum model ($\varepsilon = 0.0005$) for (a) $\gamma = \pi$ (para) and (b) $\gamma = 2\pi/3$ (meta) at very small fractions of the flux quantum $\phi_B / \phi_0 \sim 10^{-4}$.

It is interesting to compare the results shown above for the field affected electronic transmission through molecular structures to the equivalent transmission problem in the continuum ring model described in Section 2b. Fig. 12 depicts the transmission probability as function of the dimensionless parameter $kR$ calculated from Eq. (18), using the reflection parameter $\varepsilon = 0.0005$ at different flux ratios, $\phi_B / \phi_0$ of the order of $10^{-4}$. The cases $\gamma = \pi$ and $\gamma = 2\pi/3$ correspond to the para and meta connected rings, respectively. Estimating the benzene ring radius as R~0.13 nm, we find that $kR = 1$ correspond to a free electron energy of the order ~ 2eV, while $\phi_B / \phi_0 = 10^{-4}$ corresponds to $B$ of order 10 tesla. The close similarity of the results displayed in Fig. 12 and those of Fig. 8 is obvious, showing that the complex nature of the magnetic field dependent transmission probability through junctions involving rings of various geometries is intimately related to the symmetry of the wavefunctions obtained by a simple model of a particle on a ring.

This further justifies the use of the simplified two-level model introduced in Section 2c which relates the generic nature of the magnetic field dependence described above to the phase dependent coupling coefficients and magnetically controlled wave interferences on the ring. In Fig. 13, we plot the transmission probability as function of energy obtained by the two-level model expression (Eq. (26)) for two system geometries under various magnetic field intensities.

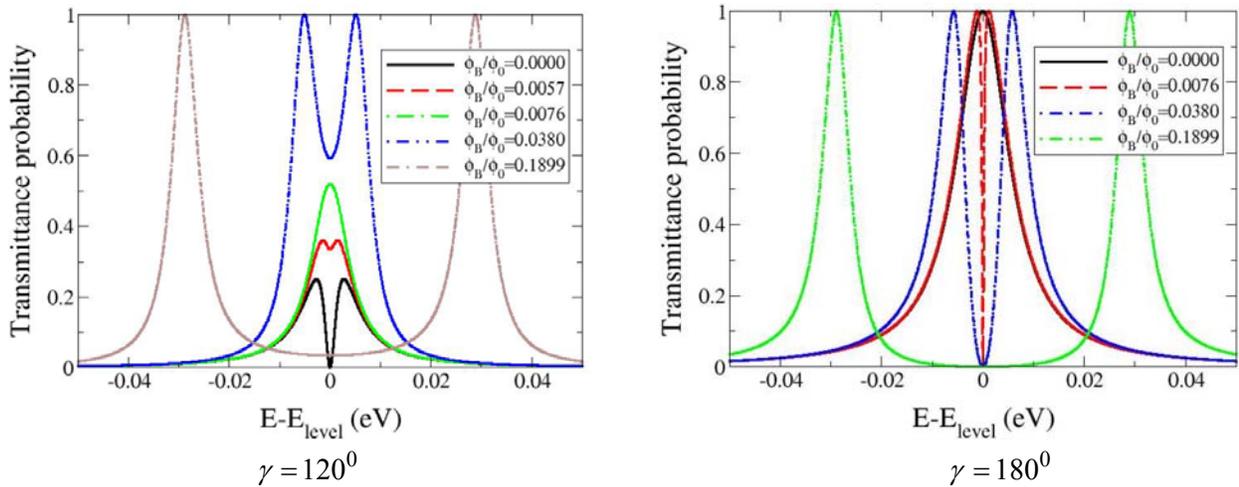

Fig. 13. Transmission probability as a function of energy obtained using the two-level model, Eq. (26), for meta (left panel) and para (right panel) connected rings. We use m=2, ring's radius= 1.0 nm, V=0.1 eV, $\rho$=0.05 eV$^{-1}$ and let the x-axis origin follow the average position of the two levels in the presence of the magnetic field. The ortho-connected ring results are identical to those obtained for the meta configuration.

As can be seen, the two-level model fully captures all the features appearing in the transmission probability curves obtained by both the tight-binding Hamiltonian and the continuum scattering description. Here, as well, for the symmetrically connected ring the magnetic field serves to split the energy levels while conserving the total area under the transmission peaks, whereas for the para and ortho configurations the integrated transmission probability grows with the magnetic field. This equivalence between the three approaches (tight binding Hamiltonian, scattering model, and two-level model) sheds light on the origin of the different behavior of the transmission probability between the three system geometries.

As suggested by the two-level model, the differences between the three geometries enter via the coupling integrals between the ring and the leads taken to be proportional to the phase of the wave function at the locations of the junctions. Coupling of one of the leads to a nodal point of the wave function will cause destructive interferences which may be lifted by the application of the external magnetic field. Since different geometries couple the leads with different phases the response of the transmission probability towards the magnetic field is altered.

## 4. Effect of dephasing

Since much of the effects discussed above result from interference between transmission pathways, it is expected that dephasing processes will have a strong effect on these observations. Such effects were studied by several authors in this context using the Büttiker probe method[56] whose application predominates the field of junction transport. Because this phenomenological method is based on a rather artificial process of replacing coherently transmitted electrons by electrons with indeterminate phase, we chose to compare such results with those obtained from another phenomenological model in which the dynamics of the density matrix of the molecular bridge incorporates damping of non-diagonal elements as done in the Bloch or Redfield equations describing relaxation in a multilevel system.

In both the Büttiker probe and the density matrix approaches it is possible to affect dephasing locally, i.e. at any given site of the tight-binding bridge that represents our molecule, using the following procedures:

In the Büttiker probe method[56, 57] the dephasing rate at such a site, $j$, is determined by its coupling to an external thermal electron reservoir, $J$, with chemical potential set such that no net current flows through the corresponding contact. We describe the probe by the same tight-binding metal model, Eq. (2), and the same energetic parameters (site energy and intersite coupling) as our source and drain leads. The dephasing rate is determined by the coupling $\beta_{j,J}$ between molecular site $j$ and the site of the probe $J$ that is coupled to it. In the calculations reported below we take all these coupling parameters between molecular sites and probes to be the same, $\beta_{j,J} = \beta_{MB}$, for all $j$ and $J$. Within the model, one calculates the transmission functions $T_{K,K'}(E)$ between any two leads ($K, K' = L, R, J = 1,...N$ for $N$ probes) and obtains the effective transmission function between source and drain in the form

$$T_{RL}^{(eff)} = T_{RL} + \sum_{J,J'=1}^{N} T_{RJ} \, W_{JJ'} \, T_{J'L} \, , \text{ where } W^{-1}{}_{JJ'} = -T_{JJ'}(1-\delta_{JJ'}) + (1-\mathcal{R}_{JJ})\delta_{JJ'} \, , \, (\delta \text{ is}$$

the Kronecker delta function) and $R_{JJ} = 1 - \sum_{K=R,L,\{J'\neq J\}} T_{JK}$ is the reflection function in probe $J$.

In the density matrix description[36, 37] one looks, for a given injection current in the source wire, for the steady state solution of the Liouville equation for the density matrix of the inner system

$$\dot{\rho}_{nm} = -\frac{i}{\hbar}[\hat{H},\hat{\rho}]_{nm} - \frac{1}{2}(\Sigma_n(E) + \Sigma_m(E)) - \frac{1}{2}\eta_{nm}(1-\delta_{nm})\rho_{nm} \, . \qquad (27)$$

where $\Sigma_n$ is the self-energy, Eq. (9), representing the effect of an infinite lead coupled to site $n$ and $\eta_{nm} = (1/2)\eta(\delta_{nM} + \delta_{mM})$, where $\delta_{nM}$ is 1 if site $n$ is on the molecule and is zero otherwise. Here, the parameter $\eta$ (taken to be the same for all ring sites) represents the dephasing rate. The resulting steady state solution is then used to evaluate the current on any bond segment as well as the transmission coefficient.[36, 37]

Figures 14 compares results from these calculations in the absence of a magnetic field. Here dephasing is affected on all sites of the molecular (benzene) ring and the transmission coefficient is plotted against electron energy for the para, meta and ortho connected benzene molecules for different values of the dephasing parameters. We note in passing that our calculations using the Büttiker probe method are practically identical to those obtained by Dey et al.[58] when the same junction parameters are used. On this level of presentation the main effect of dephasing is seen to be broadening of the transmission peaks. We note that the two different phenomenological models of dephasing give qualitatively similar results.

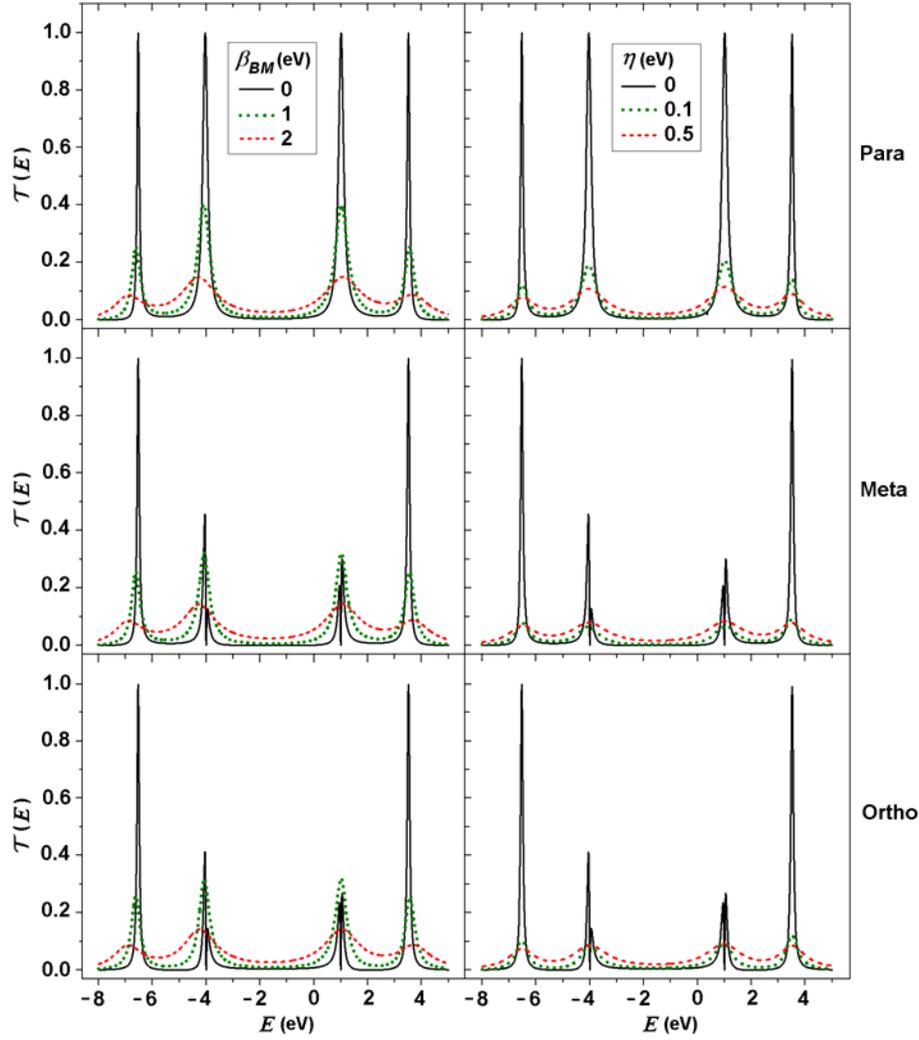

Fig. 14. Transmission probability as a function of energy in the presence of dephasing: Top, middle and bottom figures correspond to para-, meta- and ortho-connected benzene molecules. Left: Results obtained using the Büttiker probe model with the indicated coupling parameter $\beta_{BM}$. Right: results obtained by the density-matrix calculation with the indicated dephasing rate $\eta$. The molecule-leads coupling is 1 eV. The other model parameters are as given in the second paragraph of Section 3 (parameters of the probe leads are the same as for the source and drain leads).

Figures 15 (using the Büttiker probe method) and 16 (density matrix model) focus on the transmission resonance near $E = 1\,\text{eV}$. The broadening effect caused by dephasing can be clearly seen both in the strong (upper panels) and weak (middle panels) leads-ring coupling regimes. In addition the effect of dephasing on eliminating interference characteristics is apparent. This is most pronouncedly manifested in the integrated transmission (lower panels). For low bias and temperature, broadening alone makes the integral smaller at any finite $E$ because part of the integrand exits the narrow integration window. Furthermore, in the para connected molecule, the integrated transmission goes down also because of the destruction of constructive interference. Conversely, in the meta configuration it goes up (Fig. 15) or considerably more weakly down (Fig. 16) because the destructive interference is eliminated.

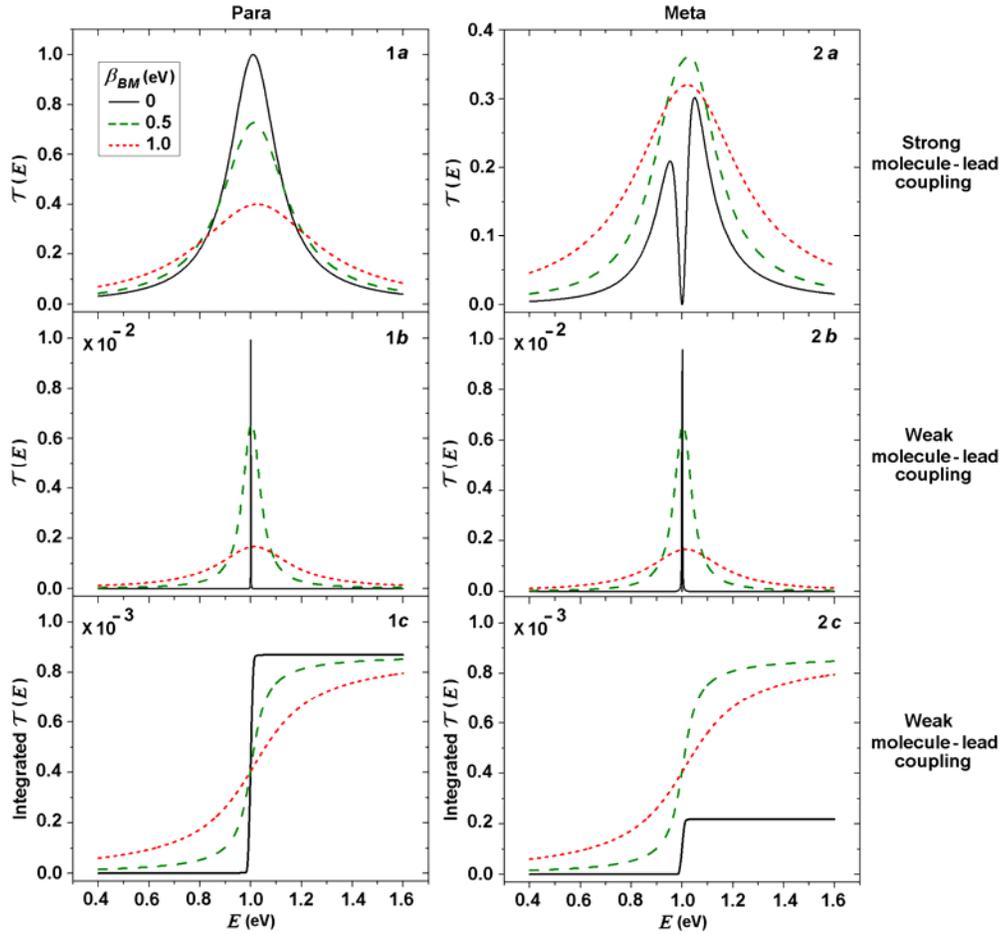

Fig. 15. Dependence of the transmission resonance at 1 eV on dephasing calculated with the Büttiker probe method. The line style and color representing different values of $\beta_{BM}$ and given in the framed inset in panel 1a correspond to all panels. Panels 1a,b,c show results for para-connected benzene while panels 2a,b,c correspond to the meta-connected molecule. In the $a$ and $b$ panels the molecule-source/drain couplings are $\beta_{KM} = 1$ eV and 0.05 eV, respectively ($K = L, R$). In the weak molecule-source/drain coupling case (panels 1b, 2b) the $\beta_{BM} = 0$ (no dephasing) lines are scaled down by a factor of 100 (i.e. multiplied by 1/100) in the para case and by a factor of 2 in the meta case, in order to fit on the given scale. The c panels show the integral about the resonance, $\int_a^E dE T(E)$ for the weak coupling (0.05 eV; as in panel s b) case plotted against $E$, where $a < 1$ $eV$ is placed well below the 1eV resonance but well above the lower transmission resonance.

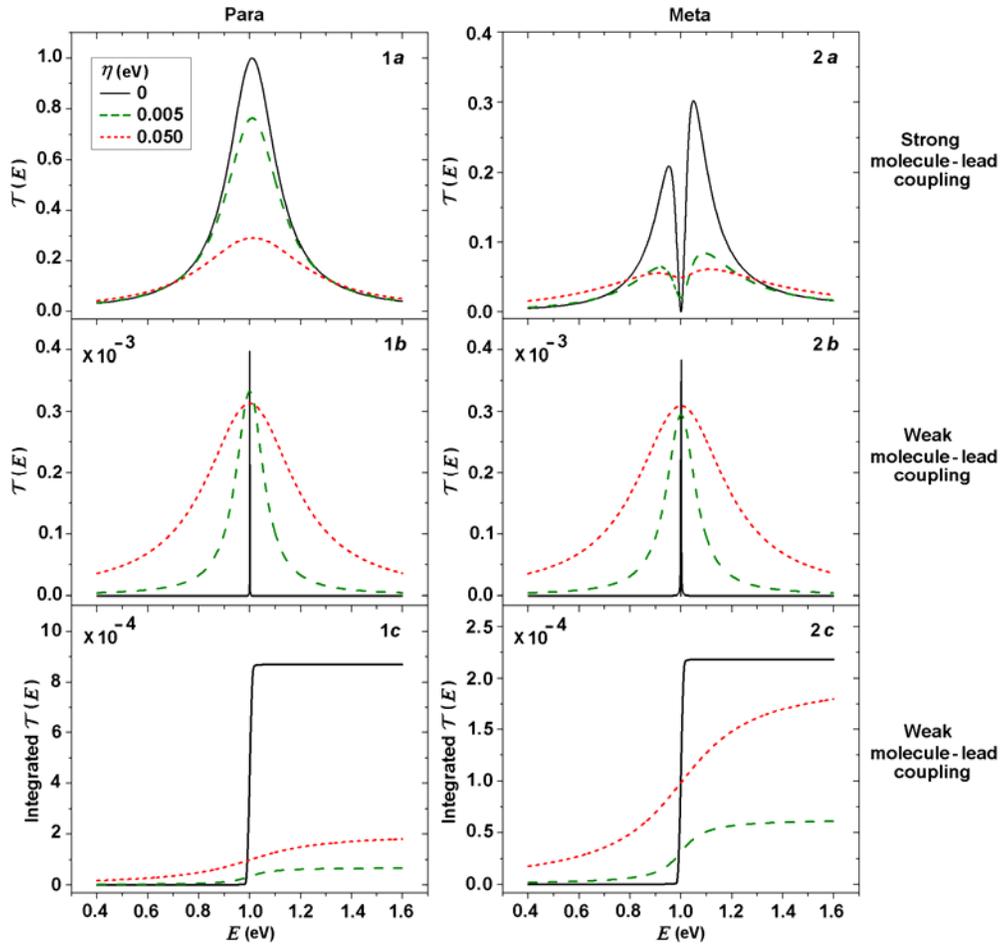

Fig. 16. Same as Fig. 15, where the effect of dephasing is obtained from the density matrix approach. The line style and color representing different values of $\eta$ and given in the framed inset in panel 1a correspond to all panels. In the *b* panels, the $\eta = 0$ lines are scaled down by the multiplicative factor of $4 \times 10^{-4}$ in the para case, and $2 \times 10^{-2}$ in the meta connected molecule in order to fit into the scale used here. Note that the vertical scale itself goes up to $4 \times 10^{-4}$, i.e. the peak transmission in the para-connected molecule without dephasing is 1.

In Figures 7 and 8 above we have demonstrated the possibility of magnetic field control of the I-V characteristic in the case of meta (and ortho) connected benzene. Figures 17-18 show the effect of dephasing on this dependence. Figures 17a is the analog of Fig. 7b, where the I-V characteristic for the meta-connected benzene under different magnetic fields normal to the ring plane is shown, now in the presence dephasing (Büttiker probe method, $\beta_{BM} = 0.1\,\text{eV}$). Similar results are obtained when dephasing is introduced in the density matrix method. We see that the field dependence of the current step strongly diminishes in the presence of dephasing. This is shown more explicitly in Figures 17b,c, where the difference between the current evaluated at $V = 2.2$ V for $B = 0$ and $B = \pm 15\,\text{tesla}$ is plotted against the dephasing parameter.

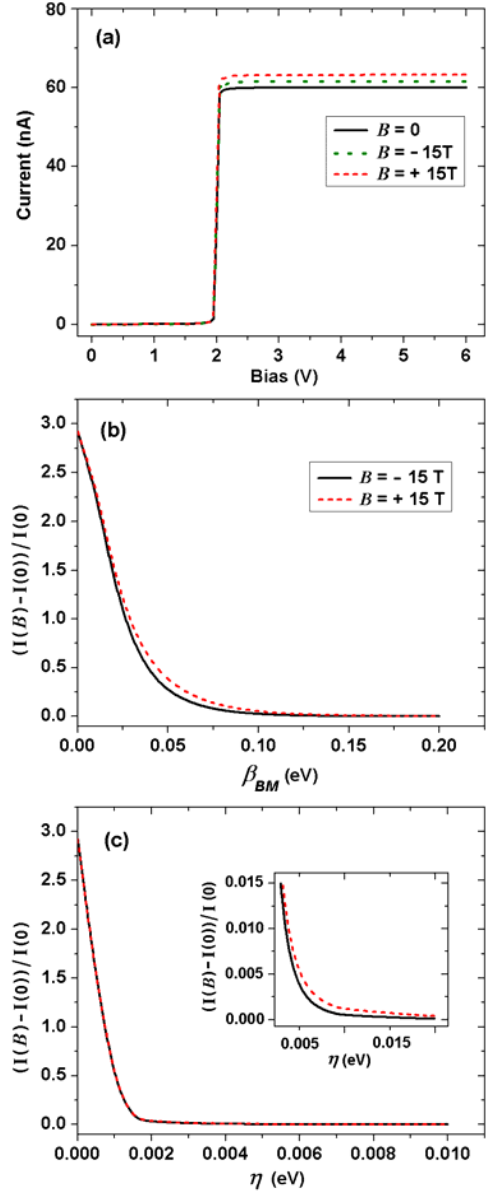

Fig. 17. (a) Magnetic field dependence of the current near the 2V step (associated with the transmission resonance at $E = 1\,\mathrm{eV}$; analog of Fig. 7b) calculated for a weakly coupled ($\beta_{KM} = 0.05\,\mathrm{eV}$) meta-connected junction with dephasing implemented by the Büttiker probe method ($\beta_{BM} = 0.1\,\mathrm{eV}$). (b) and (c) The same magnetic field dependence expressed by the ratio $[I(B)-I(0)]/I(0)$ plotted respectively against the dephasing parameters $\beta_{BM}$ (Büttiker probe method) and $\eta$ (density matrix method). The inset in 17c displays the results of the main figure on a different scale, emphasizing the observation (also seen in 17a) of deviations from $\pm B$ symmetry at intermediate dephasing rates.

More insight about the origin of this behavior is obtained from Fig. 18, which is the analog of Fig. 8 calculated in the presence of dephasing. The primary effect of dephasing in the range considered is seen to be the destruction of interference effects, which strongly diminishes the difference between the transmission properties of the para and meta connected junctions as well as the dependence on an imposed magnetic field.

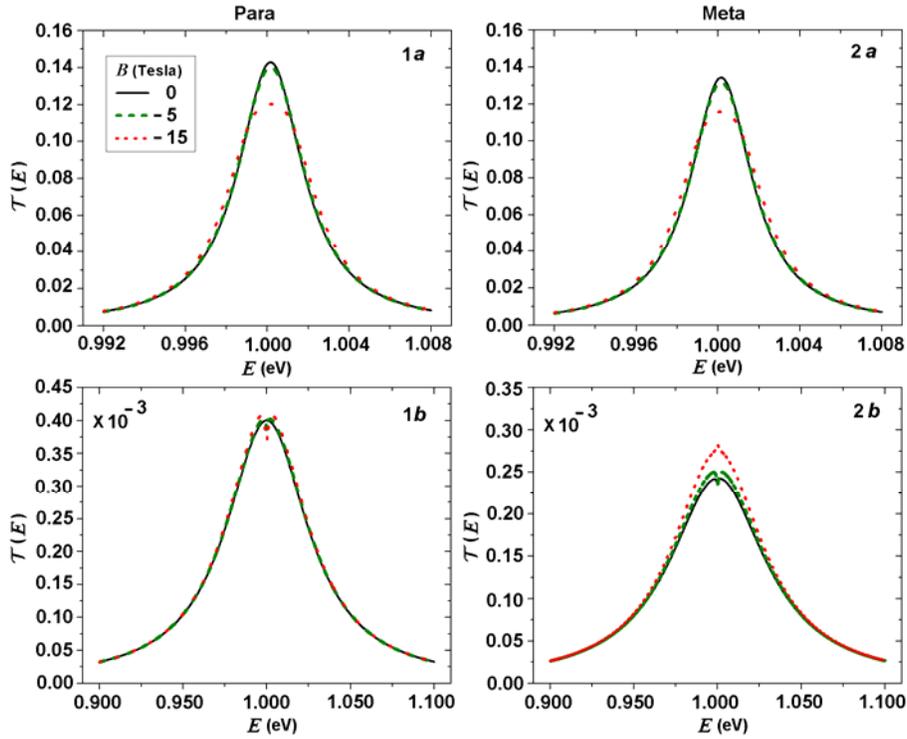

Fig. 18. Transmission probability for para (1a, b) and meta (2a, b) weakly connected ($\beta_{KM} = 0.05\,\text{eV}$) benzene as a function of electron energy in the presence of dephasing affected through Büttiker probes (panels *a*) ($\beta_{BM} = 0.1\,\text{eV}$) and density matrix method (panels *b*) ($\eta = 0.001\,\text{eV}$) at $B = 0$, -5 and -15 Tesla.

The significance of the dynamic destruction of phase in the quick erasure of the magnetic field effect on the conduction properties of asymmetrically connected benzene molecules can be gauged against other effects that can potentially affect this sensitivity. First, our calculations have disregarded the implications of the voltage distribution across the molecular junction. On the simplest level of description this can appear in the voltage division between the two contact, expressed by a factor $\xi$ and a voltage $\Delta V$ such that $\xi(V - \Delta V)$ and $(1-\xi)(V - \Delta V)$ are the potential drops on the two molecule-lead contacts while $\Delta V$ is the potential drop on the molecule itself. All the calculations described above where done with $\xi = 0.5$ and $\Delta V = 0$. We have established that our results are not sensitive to the choice of $\xi$. An example of the effect of $\Delta V$ is shown in Fig. 19, which extends the calculations displayed in Fig. 8b (meta-connected benzene) to the case shown in panel (a) where some of the site energies are changed, $\alpha \to \alpha \pm \Delta\alpha$. We see that the magnetic field effect on the transmission decreases with increasing $\Delta\alpha$ and practically disappears for $\Delta\alpha = 0.05$ eV. This emphasizes the need to carry such experiments under low bias conditions, implying that need to align the molecular spectrum with respect to the lead Fermi energy with a gate potential.

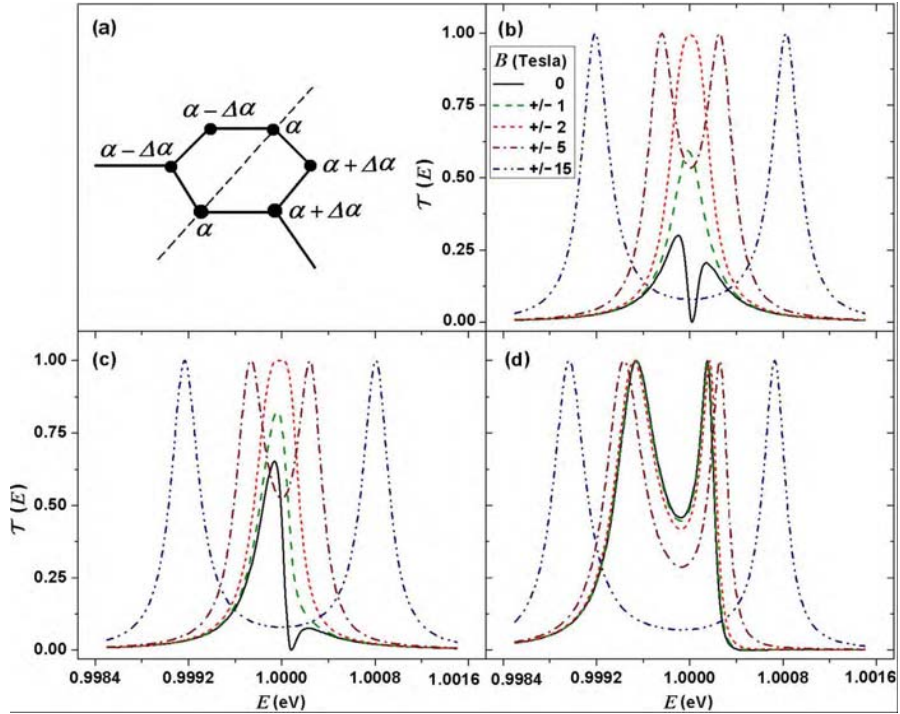

Fig. 19. The transmission function $T(E)$ for the meta-connected Benzene, same as Fig. 8b, with some site energies shifted as shown in panel (a). In panels (b), (c), (d) $\Delta\alpha$ is taken 0.01, 0.02 and 0.05 eV, respectively.

Secondly, keeping in mind that the required slow dephasing implies low temperature, it is important to note that the effect seen in Figs. 17, 18 arises from the dynamic destruction of phase, and is not reproduced merely by raising the electronic temperature of the leads. This is seen in Fig. 20, which shows the effect of leads electronic temperature on the observed magnetic field dependence of the conduction through a meta-connected benzene junction near the molecular resonance at 1 eV (bias voltage 2 V) in the absence of dephasing. Fig. 20a is similar to Fig. 7b except that the Fermi distributions in the electrodes were taken at 300K. Fig. 20b displays the current at V = 2.2 V through this junction, plotted against the electrodes temperature for $B$ = 0 and 15 tesla. Only weak electrode temperature effect is seen at the realistic temperature range considered.

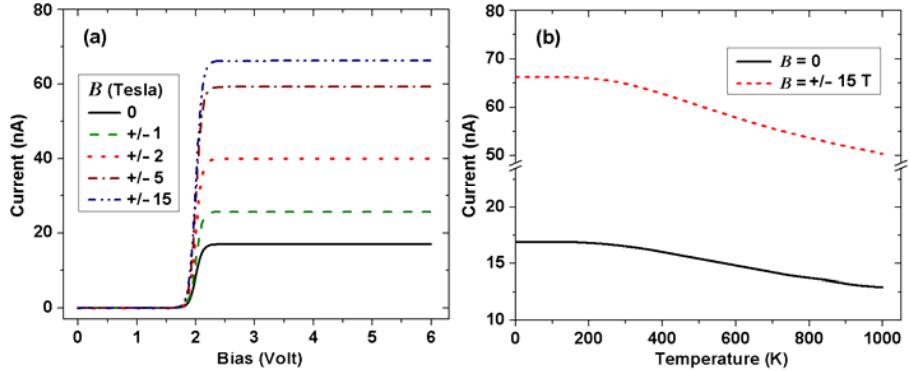

Fig. 20. (a) Current vs. bias voltage for a weakly ($\beta_{KM} = 0.05\,\text{eV}$) meta-connected benzene for different imposed magnetic fields perpendicular to the molecular plane (same as Fig. 7b) calculated at T = 300K. (b) The current through the same junction at V = 2.2 V, displayed as a function of temperature for $B$ = 0 and 15 T.

Finally, It is interesting to note that in the presence of dephasing, deviations from Onsager symmetry under reversal of field direction, $B \leftrightarrow -B$, are observed (as seen in Fig. 17). Such deviations were discussed in previous work in different contexts, including coupling to a thermal environment.[59-62] We defer further discussion of this issue to a later publication.

## 5. Concluding Remarks

In this paper we have addressed the issue of magnetic field effect on electronic transport in molecular conduction junctions. Observations of the Aharonov-Bohm effect in mesoscopic conducting loops could suggest that molecular junctions comprising molecular ring structures as bridges will show similar effects, however the small radius of molecular rings implies that the field needed to observe the AB periodicity is unrealistically large. Still, we have found that strong magnetic field effects can be seen under the following conditions: (a) The molecular resonance associated with its conduction behavior is at least doubly degenerate, as is often the case in molecular ring structures; (b) the molecule - lead coupling is weak, implying relatively distinct conduction resonances, (c) asymmetric junction structure and (d) small dephasing (implying low temperature) so as maintain coherence between multiple conduction pathways. Interestingly, in weakly connected symmetric junctions (e.g. para connected benzene) the transmission coefficient can show sensitivity to magnetic field, however it is found that the integrated transmission is field independent, so that this sensitivity is not reflected in the current-voltage characteristic.

For the organic structures we have used the within the tight-binding (Hückel) model modified for the presence of magnetic field using the London approximation. Qualitatively similar results were obtained from the analog model of a continuous ring, showing that the qualitative effect studied depends mostly on the strength and symmetry of the molecule-lead coupling. We have also shown that much of the qualitative behavior of conduction in these models can be rationalized in terms of a much simpler junction model based on a two-state molecular bridge.

When the conditions outlined above are satisfied, strong dependence of conduction on the imposed magnetic field can be found. The effect of dephasing processes on this observation are studied using two different phenomenological models: the Büttiker probe and phenomenological coherence damping imposed on the Liouville equation for the molecular density matrix. Both treatments are approximate: The approximate nature of the density matrix approach stems from the fact that dephasing was affected by damping non-diagonal elements of the density matrix in the local site representation while assuming that the transmission energy remained well defined.[36, 37] The Büttiker probe approach is limited to linear response and cannot be rigorously applied to threshold phenomena in the current-voltage dependence. Still, the fact that these two different approaches gave qualitatively similar results in all cases studied, provide some assurance about their validity. Both models show strong suppression of the sensitivity of conduction to the imposed magnetic field.

Next, consider the implications of the conditions outlined above to experimental considerations. Conditions (a) and (c) can be met by making a proper choice of molecular bridge and the positions of linker groups. Condition (b) of weak molecule-lead coupling does not imply weak molecule-lead bonding, only that the resonance states that involve multiple pathways through the ring (or counter propagating wavefunctions in the ring) are weakly coupled to the metal electrodes. This can be achieved by connecting molecular ring to leads via saturated alkane chains.

Condition (d), the requirement for small dephasing, is inherent in all experiments trying to observe interference phenomena in molecular junctions, and implies the need to work at relatively low temperatures. The Büttiker-probe procedure is not related directly to a physical process, so it is hard to assess the experimental implication of the coupling $V_{MN}$. The equivalent analysis in terms of the coherence damping rate $\eta$ does provides an estimate: For the reasonable molecular parameters chosen in our calculations, Fig. 17c shows that magnetic field effects are suppressed when this damping rate exceeds ~ 0.001 eV, that is, dephasing times of the order of 1 ps. Recent observations in different systems[63] have shown that molecular electronic coherence can persist on such timescales even at room temperatures. This suggests that the magnetic field effects discussed in this paper may be observables.

Finally, we have observed magnetic asymmetry (under reversal of field direction) in the presence of dephasing. This observation and its repercussions will be discussed elsewhere.

**Acknowledgment**

We thank Joe Imry for helpful discussions. The research of A.N. is supported by the Israel Science Foundation, the Israel-US Binational Science Foundation, the European Science Council (FP7 /ERC grant no. 226628) and the Israel – Niedersachsen Research Fund. O.H. acknowledges the support of the Israel Science Foundation under Grant No. 1313/08, the support of the Center for Nanoscience and Nanotechnology at Tel-Aviv University, and the Lise Meitner-Minerva Center for Computational Quantum Chemistry. D. R. Acknowledges a Fellowship received from the Tel Aviv University nanotechnology Center.

**Appendix A: Scattering model for asymmetric nanoscale junctions containing molecular rings and magnetic fields.**

We consider the setup presented in Fig. 2a, and write the electron wavefunction in each segment of the ring as $A_1 e^{ikl} + A_2 e^{-ikl}$ where $l$ denotes a propagation distance, i.e. $l = R\theta$ in ring segments; $\theta$ is the angle traversed by the electron, and $R$ the radius of the ring. We use the standard notation where positive $k$ represents counter-clockwise propagating waves and negative values represent clockwise moving waves. In order to calculate the transmission probability and the circular current as a function of magnetic field we assign scattering amplitudes as shown in Fig. 2b and explained in the main text. For simplicity, we assume in what follows that all scattering amplitudes are real. This assumption implies that no rigid phase shifts occur upon scattering at the junctions consistent with the tight-binding model which conserves the continuity of the wave-functions across the junctions.[42, 47, 48] Focusing on junction I (Fig. 2a) we may write the following scattering equation relating the incoming amplitudes to the outgoing amplitudes:

$$\begin{pmatrix} R_2 \\ D_2 \\ U_1 \end{pmatrix} = \begin{pmatrix} c & \sqrt{\varepsilon} & \sqrt{\varepsilon} \\ \sqrt{\varepsilon} & a & b \\ \sqrt{\varepsilon} & b & a \end{pmatrix} \begin{pmatrix} R_1 \\ D_1 \\ U_2 \end{pmatrix} \tag{A.1}$$

Current conservation implies that the scattering amplitude matrix must be unitary, i.e.

$$\begin{pmatrix} c & \sqrt{\varepsilon} & \sqrt{\varepsilon} \\ \sqrt{\varepsilon} & a & b \\ \sqrt{\varepsilon} & b & a \end{pmatrix}^2 = \hat{I} \tag{A.2}$$

where $\hat{I}$ is a $3 \times 3$ unit matrix. This provides three independent equations for the four scattering amplitudes, leading to

$$c = \sqrt{1 - 2\varepsilon} \; ; \; a = \frac{1}{2}(1 - c); \; b = -\frac{1}{2}(1 + c) \tag{A.3}$$

Having characterized the contacts we turn back to the circular setup in Fig. 2a. For a given wavenumber $\pm k$ we can write a scattering equation similar to Eq. (A.1) for contacts I and II, taking into account the spatial phase accumulated by the electrons while traveling along the arms of the ring:

$$\begin{pmatrix} R_2 \\ D_2 \\ U_1 \end{pmatrix} = \begin{pmatrix} c & \sqrt{\varepsilon} & \sqrt{\varepsilon} \\ \sqrt{\varepsilon} & a & b \\ \sqrt{\varepsilon} & b & a \end{pmatrix} \begin{pmatrix} R_1 \\ D_1 e^{i|k|R(2\pi-\gamma)} \\ U_2 e^{i|k|R\gamma} \end{pmatrix} \tag{A.4}$$

$$\begin{pmatrix} L_2 \\ D_1 \\ U_2 \end{pmatrix} = \begin{pmatrix} c & \sqrt{\varepsilon} & \sqrt{\varepsilon} \\ \sqrt{\varepsilon} & a & b \\ \sqrt{\varepsilon} & b & a \end{pmatrix} \begin{pmatrix} L_1 \\ D_2 e^{i|k|R(2\pi-\gamma)} \\ U_1 e^{i|k|R\gamma} \end{pmatrix} \tag{A.5}$$

To model the influence of an external magnetic field threading the ring we assume that the magnetic field is homogeneous and perpendicular to the plane of the ring $\vec{B} = (0,0,B_z)$ such that the vector potential may be written as:

$$\vec{A} = -\tfrac{1}{2}\vec{r}\times\vec{B} = -\frac{1}{2}\begin{vmatrix} \hat{x} & \hat{y} & \hat{z} \\ x & y & z \\ 0 & 0 & B_z \end{vmatrix} = -\frac{1}{2}(yB_z,-xB_z,0) = \frac{1}{2}B_z(-y,x,0) \tag{A.6}$$

On the circle defining the ring this becomes

$$\vec{A} = \frac{1}{2}B_z R(-\sin(\theta),\cos(\theta),0) \tag{A.7}$$

The magnetic phase accumulated by an electron traveling along the ring is thus given by:

$$\begin{aligned} \varphi_m &= -\frac{|e|}{\hbar}\int_{\theta_1}^{\theta_2} \vec{A}\cdot d\vec{l} = -\frac{1}{2}B_z R^2 \frac{|e|}{\hbar}\int_{\theta_1}^{\theta_2}(-\sin(\theta),\cos(\theta),0)\cdot(-\sin(\theta),\cos(\theta),0)d\theta = \\ &= -\frac{\phi_B}{\phi_0}(\theta_2-\theta_1) \end{aligned} \tag{A.8}$$

where $\phi_B = B_z S$, $S = \pi R^2$, and $\phi_0 = 2\pi\hbar/|e|$. In the presence of such magnetic field, Eqs. (A.4) and (A.5) are modified as follows:

$$\begin{pmatrix} R_2 \\ D_2 \\ U_1 \end{pmatrix} = \begin{pmatrix} c & \sqrt{\varepsilon} & \sqrt{\varepsilon} \\ \sqrt{\varepsilon} & a & b \\ \sqrt{\varepsilon} & b & a \end{pmatrix} \begin{pmatrix} R_1 \\ D_1 e^{i\left(|k|R-\frac{\phi_B}{\phi_0}\right)(2\pi-\gamma)} \\ U_2 e^{i\left(|k|R+\frac{\phi_B}{\phi_0}\right)\gamma} \end{pmatrix} \tag{A.9}$$

$$\begin{pmatrix} L_2 \\ D_1 \\ U_2 \end{pmatrix} = \begin{pmatrix} c & \sqrt{\varepsilon} & \sqrt{\varepsilon} \\ \sqrt{\varepsilon} & a & b \\ \sqrt{\varepsilon} & b & a \end{pmatrix} \begin{pmatrix} L_1 \\ D_2 e^{i\left(|k|R+\frac{\phi_B}{\phi_0}\right)(2\pi-\gamma)} \\ U_1 e^{i\left(|k|R-\frac{\phi_B}{\phi_0}\right)\gamma} \end{pmatrix} \tag{A.10}$$

Focusing on a scattering process with incoming electron coming from the right, we set $L_1 = 0$, that is, take zero incoming amplitude on the left lead. Eqs. (A.9)-(A.10) then lead to

$$\begin{pmatrix} 0 & be^{i\left(|k|R+\frac{\phi_B}{\phi_0}\right)(2\pi-\gamma)} & ae^{i\left(|k|R-\frac{\phi_B}{\phi_0}\right)\gamma} & -1 \\ be^{i\left(|k|R-\frac{\phi_B}{\phi_0}\right)(2\pi-\gamma)} & 0 & -1 & ae^{i\left(|k|R+\frac{\phi_B}{\phi_0}\right)\gamma} \\ ae^{i\left(|k|R-\frac{\phi_B}{\phi_0}\right)(2\pi-\gamma)} & -1 & 0 & be^{i\left(|k|R+\frac{\phi_B}{\phi_0}\right)\gamma} \\ -1 & ae^{i\left(|k|R+\frac{\phi_B}{\phi_0}\right)(2\pi-\gamma)} & be^{i\left(|k|R-\frac{\phi_B}{\phi_0}\right)\gamma} & 0 \end{pmatrix} \begin{pmatrix} D_1 \\ D_2 \\ U_1 \\ U_2 \end{pmatrix} = \begin{pmatrix} 0 \\ -\sqrt{\varepsilon}R_1 \\ -\sqrt{\varepsilon}R_1 \\ 0 \end{pmatrix} \quad \text{(A.11)}$$

Inverting (A.11) yields the wavefunctions amplitudes $D$ and $U$ on the ring segments as a function of the incoming amplitude $R_1$. In particular, the results for $D_2$ and $U_1$ can be used in (A.10) to yield $L_2$ and therefore the transmission probability

$$T(k,\phi_B) = \left|\frac{L_2}{R_1}\right|^2 = \frac{\varepsilon}{|R_1|^2}\left|D_2 e^{i\left(|k|R+\frac{\phi_B}{\phi_0}\right)(2\pi-\gamma)} + U_1 e^{i\left(|k|R-\frac{\phi_B}{\phi_0}\right)\gamma}\right|^2 \quad \text{(A.12)}$$

which finally results in Eq. (18)

**Appendix B: The two-level transport model**

Here we construct a two-level transport model that, for weak molecule-lead coupling, captures the main features observed in the magnetic field dependence of electron transmission through a molecular ring. The validity of a two-level model stems for the fact that in this coupling limit only pairs of molecular levels, degenerate in the limit of zero coupling, are coupled through their mutual interaction with the leads. The Hamiltonian for this two-level system can be obtained by considering the Hamiltonian, $\hat{H} = (1/2)\left(\vec{\hat{P}} - q\vec{\hat{A}}\right)^2 + \hat{V}(\vec{r})$ for a charged particle moving in a magnetic field. The Hamiltonian describing a free ($V=0$) electron ($q = -|e| = -1$; atomic units are used throughout) moving on a circular ring of radius R lying in the XY plane under a uniform magnetic field oriented in the Z direction, $B = (0,0,B_z)$ can be written by setting

$$\vec{\hat{A}} = -\frac{1}{2}\vec{r}\times\vec{B} = \frac{B_z}{2}(-y,x,0) \quad \text{(B.1)}$$

This leads to

$$\hat{H} = \frac{1}{2}\left[\vec{P}^2 + \vec{A}^2 + \vec{A}\cdot\vec{P} + \vec{P}\cdot\vec{A}\right] = -\frac{1}{2}\nabla^2 - i\frac{B_z}{2}\left[x\frac{\partial}{\partial y} - y\frac{\partial}{\partial x}\right] + \frac{B_z^2}{8}\left(x^2 + y^2\right) \quad \text{(B.2)}$$

where the central term in the last stage may be identified as the angular momentum component along the Z direction. In circular coordinates with the origin placed at the center of the ring Eq. (B.2) becomes

$$\hat{H} = -\frac{1}{2R^2}\frac{\partial^2}{\partial\theta^2} - i\frac{B_z}{2}\frac{\partial}{\partial\theta} + \frac{R^2 B_z^2}{8} \quad \text{(B.3)}$$

in which the last term is an additive constant. The eigenstates of this Hamiltonian can be written in the form

$$\psi(\theta) = \frac{1}{\sqrt{2\pi}}e^{im\theta}; \quad m = 0, \pm 1, \pm 2, \cdots \quad \text{(B.4)}$$

with the corresponding energy eigenvalues

$$E_m = \frac{1}{2R^2}\left(m + \frac{\phi_B}{\phi_0}\right)^2 \quad \text{(B.5)}$$

Here, $\phi_B = \vec{B}\cdot\vec{S}$ and $\phi_0 = 2\pi$ are the magnetic flux threading the ring and the flux quantum, respectively. Using relation (B.4) the wave functions can be rewritten as

$$\psi_{cw}(\theta) = \frac{1}{\sqrt{2\pi}}e^{-i\left(\sqrt{2E_m}R + \frac{\phi_B}{\phi_0}\right)\theta}; \quad \psi_{ccw}(\theta) = \frac{1}{\sqrt{2\pi}}e^{i\left(\sqrt{2E_m}R - \frac{\phi_B}{\phi_0}\right)\theta}, \quad \text{(B.6)}$$

Noting that Eq. (B.5) may be written in the form

$$E_{\pm|m|} \equiv \frac{1}{2R^2}\left[m^2 + \left(\frac{\phi_B}{\phi_0}\right)^2\right] \pm \frac{|m|B_z}{2}, \quad \text{(B.7)}$$

the Hamiltonian of the isolated ring in the subspace of these two levels is given by Eq. (20). When coupled to leads as in Fig. 3, the self-energy terms appearing in Eq. (23c) are given by

$$\Sigma_{L/R}^r(E) = \begin{pmatrix} -V_{1,L/R}^* & 0 & 0 & \cdots \\ -V_{2,L/R}^* & 0 & 0 & \cdots \end{pmatrix} \begin{pmatrix} \left[G_{L/R}^{r0}(E)\right]_{11} & \left[G_{L/R}^{r0}(E)\right]_{12} & \cdots \\ \left[G_{L/R}^{r0}(E)\right]_{21} & \left[G_{L/R}^{r0}(E)\right]_{22} & \cdots \\ \vdots & \vdots & \ddots \end{pmatrix} \begin{pmatrix} -V_{1,L/R} & -V_{2,L/R} \\ 0 & 0 \\ 0 & 0 \\ \vdots & \vdots \end{pmatrix} =$$

$$= \left[G_{L/R}^{r0}(E)\right]_{11} \begin{pmatrix} |V_{1,L/R}|^2 & V_{1,L/R}^* V_{2,L/R} \\ V_{2,L/R}^* V_{1,L/R} & |V_{2,L/R}|^2 \end{pmatrix}$$

(B.8)

Here, $G_{L/R}^{r0}(E)$ is the retarded (r) Green's function of the isolated (0) left/right (L/R) lead and we assume short range interaction between the ring and the leads, whereby the ring is coupled to the nearest neighbor leads sites. The advanced self-energy is given by $\Sigma_{L/R}^{a}(E) = \left[\Sigma_{L/R}^{r}(E)\right]^{\dagger}$.

The coupling matrix elements $V_{j,K}$; $j=1,2$; $K=L,R$ that appear in expression (B.8) for the self-energy are formally calculated via $\langle \psi_{ring} | \hat{V} | \psi_{lead} \rangle$ and should therefore be proportional to the phase of the wave function on the ring. In Eq. (24) we take this phase dependence into account where the specific symmetry of the system (ortho, meta, or para) is taken explicitly into account via the angular separation between the leads. Using the coupling matrix elements given in Eq. (24) we obtain explicit matrix representations for the retarded (and advanced) self-energies in the forms

$$\Sigma_R^r(E) = -i\pi\rho V^2 \begin{pmatrix} 1 & 1 \\ 1 & 1 \end{pmatrix}; \quad \Sigma_L^r(E) = -i\pi\rho V^2 \begin{pmatrix} 1 & e^{2im\gamma} \\ e^{-2im\gamma} & 1 \end{pmatrix} \quad (B.9)$$

And $\Sigma^a(E) = \left[\Sigma^r(E)\right]^{\dagger}$, i.e.

$$\Sigma_R^a(E) = i\pi\rho V^2 \begin{pmatrix} 1 & 1 \\ 1 & 1 \end{pmatrix}; \quad \Sigma_L^a(E) = i\pi\rho V^2 \begin{pmatrix} 1 & e^{2im\gamma} \\ e^{-2im\gamma} & 1 \end{pmatrix} \quad (B.10)$$

The broadening matrices $\Gamma$, Eq. (23a), and the ring Green's functions, Eq. (23b) are then obtained in the forms

$$\Gamma_R(E) = 2\pi\rho V^2 \begin{pmatrix} 1 & 1 \\ 1 & 1 \end{pmatrix} \ ; \quad \Gamma_L(E) = 2\pi\rho V^2 \begin{pmatrix} 1 & e^{2im\gamma} \\ e^{-2im\gamma} & 1 \end{pmatrix} \quad (B.11)$$

$$G_M^r(E) = \left[EI - H_M^m - \Sigma_L^r(E) - \Sigma_R^r(E)\right]^{-1}$$
$$= \left[\begin{pmatrix} E - E_1 + 2i\pi\rho V^2 & 2i\pi\rho V^2 e^{im\gamma}\cos(m\gamma) \\ 2i\pi\rho V^2 e^{-im\gamma}\cos(m\gamma) & E - E_2 + 2i\pi\rho V^2 \end{pmatrix}\right]^{-1} \quad (B.12)$$

where we have used Eq. (20) for the Hamiltonian of the isolated ring. Inverting the matrix in (B.12) and in the corresponding expression for $G_M^a(E)$ leads to

$$G_M^r(E) = \frac{1}{\left(E - E_1 + 2i\pi\rho V^2\right)\left(E - E_2 + 2i\pi\rho V^2\right) + 4\pi^2\rho^2 V^4 \cos^2(m\gamma)} \times$$

$$\begin{pmatrix} E - E_2 + 2i\pi\rho V^2 & -2i\pi\rho V^2 e^{im\gamma}\cos(m\gamma) \\ -2i\pi\rho V^2 e^{-im\gamma}\cos(m\gamma) & E - E_1 + 2i\pi\rho V^2 \end{pmatrix} \quad (B.13)$$

And the retarded counterpart is

$$G_M^a(E) = \left[G_M^r(E)\right]^\dagger = \frac{1}{\left(E - E_1 - 2i\pi\rho V^2\right)\left(E - E_2 - 2i\pi\rho V^2\right) + 4\pi^2\rho^2 V^4 \cos^2(m\gamma)} \times$$

$$\begin{pmatrix} E - E_2 - 2i\pi\rho V^2 & 2i\pi\rho V^2 e^{im\gamma}\cos(m\gamma) \\ 2i\pi\rho V^2 e^{-im\gamma}\cos(m\gamma) & E - E_1 - 2i\pi\rho V^2 \end{pmatrix}$$

(B.14)

With these explicit expression for the broadening and Green's functions matrix representations, evaluating the transmission coefficient (22) becomes a lengthy but straightforward calculation leading to the final result (26).

51  the actual parameters vary slightly for the different molecules used in our calculations. We have checked that these variations do not affect our qualitative observations.

52  In the para case, one of the levels actually remains at that energy.